\documentclass[useAMS,usenatbib]{mn2e}

\usepackage{graphicx}
\usepackage{amsmath}
\usepackage{amssymb}

\newcommand{\ion}[2]{\mbox{#1$\,$\textsc{#2}}}

\title[Relics of structure formation: HVCs around M31]{Relics of structure formation: extra-planar gas and high-velocity clouds around the Andromeda Galaxy}
\author[T.~Westmeier, C.~Br\"uns, and J.~Kerp]{T.~Westmeier$^{1,2}$, C.~Br\"uns$^{2}$, and J.~Kerp$^{2}$\\
$^{1}$CSIRO Australia Telescope National Facility, PO Box~76, Epping NSW~1710, Australia\\
$^{2}$Argelander-Institut f\"ur Astronomie, Universit\"at Bonn, Auf dem H\"ugel~71, 53121~Bonn, Germany}

\begin{document}
  
  \date{Accepted 1988 December 15. Received 1988 December 14; in original form 1988 October 11}
  
  \pagerange{\pageref{firstpage}--\pageref{lastpage}} \pubyear{2008}
  
  \maketitle
  
  \label{firstpage}
  
  \begin{abstract}
    Using the 100-m radio telescope at Effelsberg, we mapped a large area around the Andromeda Galaxy in the 21-cm line emission of neutral hydrogen to search for high-velocity clouds (HVCs) out to large projected distances in excess of $100~\mathrm{kpc}$. Our $3 \, \sigma$ \ion{H}{i} mass sensitivity for the warm neutral medium is $8 \times 10^{4}~{\rm M}_{\odot}$. We can confirm the existence of a population of HVCs with typical \ion{H}{i} masses of a few times $10^{5}~{\rm M}_{\odot}$ near the disc of M31. However, we did not detect any HVCs beyond a projected distance of about $50~\mathrm{kpc}$ from M31, suggesting that HVCs are generally found in proximity of large spiral galaxies at typical distances of a few $10~\mathrm{kpc}$.
    
    Comparison with CDM-based models and simulations suggests that only a few of the detected HVCs could be associated with primordial dark-matter satellites, whereas others are most likely the result of tidal stripping. The lack of clouds beyond a projected distance of $50~\mathrm{kpc}$ from M31 is also in conflict with the predictions of recent CDM structure formation simulations. A possible solution to this problem could be ionisation of the HVCs as a result of decreasing pressure of the ambient coronal gas at larger distances from M31. A consequence of this scenario would be the presence of hundreds of mainly ionised or pure dark-matter satellites around large spiral galaxies like the Milky Way and M31.
  \end{abstract}
  
  
  \section{Introduction}
  
  One of the currently most favoured cosmological models is the so-called Lambda Cold Dark Matter ($\Lambda$CDM) model. It assumes that the evolution of the universe is dominated by dark energy and dark matter which in sum account for 96~per~cent of the total energy density of the universe \citep{Spergel2003}. An important prediction of CDM models is the hierarchical formation of gravitationally bound structures. The smallest dark-matter haloes are expected to form first, whereas larger structures, ranging from spiral galaxies to galaxy clusters, are formed at a later stage through merging and accretion of smaller dark-matter haloes (bottom-up scenario). Numerical simulations of structure formation in CDM cosmologies have successfully reproduced the mass function and radial distribution of galaxies on the scales of galaxy clusters. On smaller scales, however, simulations predict significantly more dark-matter haloes than being observed \citep{Klypin1999,Moore1999}. This discrepancy has been named the `missing satellites' problem.
  
  To overcome this problem, \citet{Blitz1999} suggested that high-velocity clouds (HVCs) might be the gaseous counterparts of the `missing' dark-matter haloes around the Milky Way. HVCs are gas clouds observed all over the sky in the 21-cm line emission of neutral atomic hydrogen. They were discovered with the Dwingeloo radio telescope by \citet{Muller1963}, and they are characterised by high radial velocities of typically $|v_{\rm LSR}| \gtrsim 100~\mathrm{km \, s}^{-1}$ (see \citealt{Wakker1991a} for details). If HVCs were the `missing' satellites, they could not have experienced significant star formation during their evolution. Therefore, they would appear in the form of pure gas clouds without any noticeable stellar population. In addition, HVCs would be spread all over the Local Group with typical distances of hundreds of kpc and high \ion{H}{i} masses of about $10^7~{\rm M}_{\odot}$.
  
  At the same time, the expected large distances of HVCs from the Milky Way would result in fairly decent angular diameters of the clouds. Therefore, \citet{Braun1999} defined a sub-sample of compact and isolated HVCs (the so-called CHVCs) which are characterised by angular sizes of less than $2^{\circ}$~FWHM as well as isolation and separation from neighbouring \ion{H}{i} emission. The overall kinematics of the CHVC population is consistent with a distribution throughout the Local Group \citep{deHeij2002b}, making them promising candidates for the `missing' dark-matter satellites around the Milky Way and the Andromeda Galaxy. \citet{Putman2002} extended the CHVC catalogue into the southern hemisphere, using the \ion{H}{i} Parkes All-Sky Survey \citep{Barnes2001}. The data for the northern and southern hemispheres were later combined by \citet{deHeij2002b} into an all-sky catalogue of 216~CHVCs.
  
  Several arguments have been raised against the idea of HVCs and CHVCs being the `missing' dark-matter haloes predicted by CDM cosmologies. First of all, attempts to identify a population of HVCs in other nearby galaxy groups \citep{Zwaan2001,Braun2001,Pisano2004} have failed, resulting in upper distance limits for HVCs and CHVCs from the Milky Way of the order of $150~\mathrm{kpc}$. Additional evidence for HVCs being nearby at distances of the order of only $10~\mathrm{kpc}$ has been provided by the detection of H$\alpha$ emission from both HVCs and CHVCs (e.g., \citealt{Kutyrev1986,Weiner2001,Tufte2002,Putman2003}) and from the determination of distance brackets for several HVC complexes (e.g., \citealt{Danly1993,Wakker1996,Wakker2007a,Wakker2007b,Thom2006,Thom2008}). Small distances from the Milky Way are also consistent with the head-tail structures found in numerous HVCs and CHVCs (e.g., \citealt{Bruens2000,Bruens2001,Westmeier2005b}) which are thought to result from ram-pressure interaction of the clouds with the ambient gas of the Galactic corona \citep{Quilis2001,Konz2002}.
  
  \begin{table}
    \caption{Observational parameters of the northern and southern part of our Effelsberg \ion{H}{i} blind survey of M31. The final baseline RMS is given for a system temperature of $25~\mathrm{K}$ at the given velocity resolution. The specified sensitivity is the baseline RMS times the spectral bin width at the original velocity resolution, converted to \ion{H}{i} column density. Sensitivities for the WNM were calculated for a brightness temperature of $T_{\rm B} > 3 \, \sigma$, a velocity resolution of $20 \; \mathrm{km \, s}^{-1}$, and a line width of $25 \; \mathrm{km \, s}^{-1}$~FWHM.}
    \label{tab_obspar}
    \begin{center}
      \begin{tabular}{lrrl}
        \hline
        parameter                 &  south &   north & unit                          \\
        \hline
        autocorrelator            &    old & 	 new &  			     \\
        bandwidth                 &    6.3 & 	  10 & MHz			     \\
        polarisations             & 	 2 & 	   2 &  			     \\
        channels per polarisation &    512 & 	4096 &  			     \\
        velocity resolution       &    2.6 & 	 0.5 & $\mathrm{km \, s}^{-1}$       \\
        frequency switching       & normal & in-band &  			     \\
        total integration time    &    180 & 	 180 & s			     \\
        final baseline RMS        & 	45 & 	  60 & mK			     \\
        sensitivity               & 	21 & 	 5.5 & $10^{16} \; \mathrm{cm}^{-2}$ \\
        WNM sensitivity           &    2.2 & 	 1.3 & $10^{18} \; \mathrm{cm}^{-2}$ \\
        WNM mass sensitivity      & 	 8 & 	   5 & $10^4 \; {\rm M}_{\odot}$	     \\
        \hline
      \end{tabular}
    \end{center}
  \end{table}
  
  The problem of determining the spatial distribution of HVCs can ultimately be solved by searching for the expected HVC population around the nearest large spiral galaxy, the Andromeda Galaxy (M31). First, the distance of M31 is well known so that important physical parameters of HVCs, such as their \ion{H}{i} mass or their diameter, can directly be determined from the observations. Second, we will look at the HVC population of M31 from the outside, allowing us to determine the (projected) radial distribution of HVCs. Finally, M31 is relatively close to the Milky Way. Therefore, the sensitivity and angular resolution of large single-dish radio telescopes will easily allow us to detect HVCs of the mass and size of the large HVC complexes observed near the Milky Way. The first comprehensive search for HVCs around M31 was carried out by \citet{Thilker2004} with the 100-m Green Bank Telescope. They mapped an area of $7^{\circ} \times 7^{\circ}$ in the 21-cm line of \ion{H}{i} with high sensitivity and discovered a population of about 20~HVCs out to the edge of their map at about $50~\mathrm{kpc}$ projected distance from M31. Their discovery marked the first detection of an extensive HVC population around a galaxy other than our own. Several of these HVCs were studied in detail with the Westerbork Synthesis Radio Telescope (WSRT) by \citet{Westmeier2005a}.
  
  Thus, the radial extent of the population of HVCs and CHVCs around galaxies like the Milky Way and M31 is confined by two limits. An upper limit of about $150~\mathrm{kpc}$ can be derived from the non-detection of HVCs in other galaxy groups, and a lower limit of about $50~\mathrm{kpc}$ is marked by the edge of the \ion{H}{i} survey of \citet{Thilker2004} out to which HVCs were found to be present. Therefore, we decided to carry out a complementary \ion{H}{i} survey with the 100-m radio telescope at Effelsberg to search for HVCs and CHVCs around M31 out to much larger projected distances in excess of $100~\mathrm{kpc}$. This would allow us to trace the distribution of the HVC population of M31 over its entire radial extent and compare our results with the predictions of CDM structure formation scenarios.
  
  Our paper is organized as follows. In Sect.~\ref{sect_observations} we explain the technical aspects of our observations. In Sect.~\ref{sect_reduction} our data reduction, calibration, and analysis strategy is described. In Sect.~\ref{sect_completeness} we discuss the various completeness issues of our survey and their implications for our results. In Sect.~\ref{sect_results} we describe the derived observational and physical properties of the high-velocity clouds detected in our survey. In Sect.~\ref{sect_comparison} we discuss the results of our comparison of the observational parameters of the HVCs near M31 with different CDM-based models and with the distribution of satellite galaxies around M31. Sect.~\ref{sect_origin} discusses the evidence for different hypotheses on the origin of HVCs. Finally, Sect.~\ref{sect_summary} summarises our results and conclusions.

  \begin{figure*}
    \includegraphics[width=0.74\linewidth]{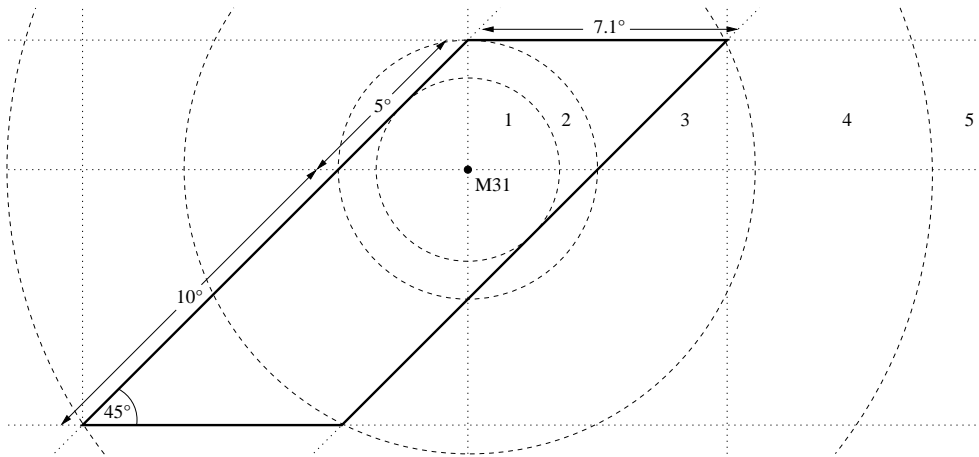}
    \caption{The plot shows the geometry of the field (thick solid line) mapped around M31. The dashed circles separate the five different regions (labelled with numbers) for which individual spatial completeness calculations have to be performed.}
    \label{fig_spatcomp}
  \end{figure*}

  \section{Observations}
  \label{sect_observations}
  
  The observations for our \ion{H}{i} blind survey of M31 were carried out between July 2003 and August 2004 with the 100-m radio telescope at Effelsberg. The observational parameters of the survey are summarised in Table~\ref{tab_obspar}. As a compromise between spatial coverage and observing time we chose the trapezoidal mapping area outlined in Fig.~\ref{fig_spatcomp}. It extends out to an angular distance of about $10^{\circ}$ from M31 in the south-eastern direction, corresponding to a projected distance of about $140~\mathrm{kpc}$ which is about two thirds of the projected distance towards M33. In this direction we expect the least confusion with foreground \ion{H}{i} emission from the Milky Way. In the north-western direction our map extends out to about $5^{\circ}$, corresponding to $70~\mathrm{kpc}$ in projection.
  
  The map was split into individual rows of constant declination with 35~pointings each. The individual pointings were separated in right ascension by $760~\mathrm{arcsec} \approx \mathrm{HPBW} \times \sqrt{2}$. Adjacent rows were separated in declination by $380~\mathrm{arcsec} \approx \mathrm{HPBW} / \sqrt{2}$ and shifted in right ascension by the same amount, resulting in an orientation angle of the entire map of $45^{\circ}$ with respect to the equatorial coordinate system. As a consequence, the final map is beam-by-beam sampled and oriented almost perpendicular to the major axis of the disc of M31. Aligning the map along the minor axis of M31 has the advantage of avoiding confusion with disc emission along the major axis as far as possible.
  
  The southern part of the map was observed with the old 1024-channel autocorrelator, whereas for the northern part we used the new 8192-channel autocorrelator (AK90). The reason for changing from the old to the new autocorrelator in early 2004 was that in August~2003 the quality and stability of the spectral baselines suddenly degraded. Several maxima and minima occurred over the bandpass, requiring a higher-order polynomial to approximate the shape of the baseline. The more severe problem was that the positions and amplitudes of the maxima and minima changed significantly over short time scales of the order of one hour, making baseline correction difficult. As the quality of the spectra was not sufficient for the objectives of our M31 survey we decided to carry out test observations with the new AK90. The test observations demonstrated that the AK90 provided significantly better baseline quality and stability than the old 1024-channel autocorrelator. Therefore, we decided to continue the survey in the northern direction using the AK90 instead of the old autocorrelator. Another advantage of the AK90 is its larger bandwidth of $10~\mathrm{MHz}$. This allowed us to use the in-band frequency switching method which spends 100~per~cent of the integration time on source. As a result the baseline RMS in the northern part of the map is by a factor of about $\sqrt{2}$ lower than in the southern part. Throughout this paper we follow the conservative approach of using the 1024-channel autocorrelator specifications as the basis for all sensitivity-related parameters.

  \section{Data reduction and analysis}
  \label{sect_reduction}
  
  \subsection{Flux calibration}
  
  The standard flux calibration source S7 was observed every 6~hours and at the beginning and end of each observing run. The corresponding flux calibration factors for the two polarisations were determined by the software \textsc{nautocal} (based on the results obtained by \citealt{Kalberla1982}), yielding statistical calibration errors of the order of only 1~per~cent over an entire observing run of typically 10 to 15~hours.
  
  \subsection{Stray radiation}
  
  A potential problem could be that \ion{H}{i} emission from the disc of M31 could have been detected through the near side lobes of the telescope and create the impression of extra-planar gas in proximity of the disc. The beam pattern of the Effelsberg telescope at $\lambda = 21~\mathrm{cm}$ was studied in detail by \citet{Kalberla1980} through combined observations of Cygnus~A and Cassiopeia~A. The overall crosswise structure in their fig.~1 results from diffraction by the four support legs. Significant side lobes can be found particularly in the north-south direction. At $1^{\circ}$ angular distance from the pointing direction the sensitivity drops to about $-30~\mathrm{dB}$. Typical brightness temperatures of $20~\mathrm{K}$, as observed along the \ion{H}{i} ring of M31, would therefore be attenuated to about $20~\mathrm{mK}$ which is far below the noise level of our data at the original spectral resolution. The near side lobes in the east-west direction are not as extended as those in the north-south direction, but instead the sensitivity levels are somewhat higher near the pointing centre. At an angular distance of $15~\mathrm{arcmin}$ the sensitivity drops to about $-20~\mathrm{dB}$, resulting in an attenuation of a $20~\mathrm{K}$ signal to about $200~\mathrm{mK}$. This is of the same order of magnitude as the expected signals of high-velocity clouds around M31. The near side lobes, however, will only affect regions of the map which are in projection close to the \ion{H}{i} disc of M31. All areas beyond a projected distance of the order of $10~\mathrm{kpc}$ from the disc will not be influenced anymore by stray radiation from M31. In addition, stray radiation through the near side lobes should result in more extended emission, whereas most HVCs and in particular CHVCs are expected to be unresolved. Consequently, stray radiation will not affect our observations since most of the expected HVCs and CHVCs should be located at sufficiently large angular distances from the disc of M31 and will likely be unresolved by the HPBW of~$9~\mathrm{arcmin}$.
  
  \subsection{Spectral baseline correction}
  
  To improve the quality of the spectral baseline we implemented a two-step baseline correction procedure. The first step was applied to each of the two polarisations separately. All 35~spectra in a single row of the map were averaged, and a polynomial of usually \mbox{8$^{\rm th}$ order} over $\Delta v = 900~\mathrm{km \, s}^{-1}$ was fitted to the averaged spectrum after manually setting line windows. This constant reference polynomial -- representing the average baseline over a time period of about two hours -- was then subtracted from each individual spectrum of the row. The resulting spectra were already quite smooth as the baseline shape at Effelsberg was fairly stable over a period of two hours. The second step was to average all spectra at a single position on the sky and fit a polynomial to the resulting spectrum. Because of the previous subtraction of the reference polynomial, a low-order polynomial was sufficient for the final averaged spectrum. In most cases, the baseline shape could be sufficiently described by a \mbox{4$^{\rm th}$} or \mbox{5$^{\rm th}$} order polynomial across a velocity range of $900~\mathrm{km \, s}^{-1}$. Only in very few cases a lower (down to \mbox{1$^{\rm st}$}) or higher (up to \mbox{8$^{\rm th}$}) order was necessary. After cutting out a few strong radio frequency interference signals by hand, the resulting spectra were Hanning-smoothed to $4.1~\mathrm{km \, s}^{-1}$ velocity resolution and combined into a FITS data cube for further analysis.
  
  \subsection{Identification of sources}
  
  The final survey consists of 3586~individual pointings for each of which an \ion{H}{i} spectrum in the velocity range of $v_{\rm LSR} \approx {-750} \ldots {+150}~\mathrm{km \, s}^{-1}$ was available for further analysis. As we were interested in the detection of HVCs and CHVCs around M31, the survey was particularly optimised for the detection of compact sources with narrow spectral lines. In the in-band frequency switching mode nearly 100~per~cent of the integration time is spent on source, maximising the sensitivity for the detection of faint CHVCs. The spectral baselines in this special observing mode, however, are not particularly smooth and stable, restricting the detectability of faint, diffuse sources.
  
  \begin{figure}
    \centering
    \includegraphics[width=0.912\linewidth]{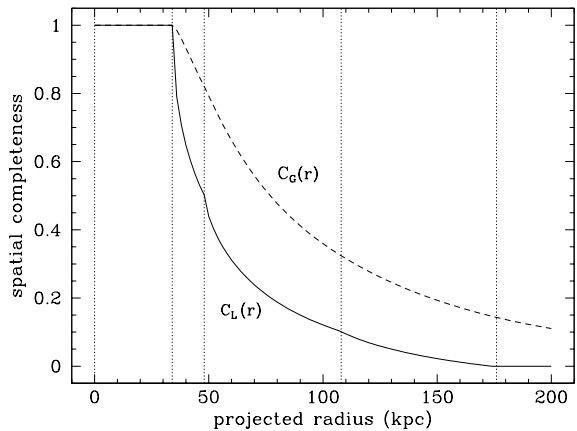}
    \caption{Spatial completeness of the M31 survey as a function of projected distance, $r$, from the centre of M31. The dashed line shows the global completeness function, $C_{\rm G}(r)$, and the solid line the local completeness function, $C_{\rm L}(r)$. The vertical dotted lines indicate the radii $a$, $b$, $c$, and $d$ at which the completeness function changes abruptly due to the special geometry of the observed field.}
    \label{fig_spatcomp2}
  \end{figure}
  
  To identify HVC/CHVC candidates of M31 all spectra were searched by eye for potential emission lines. The criterion for HVC emission was a separation of the signal in phase space from the disc emission of M31 (and the Milky Way), where phase space is considered the observable three-dimensional sub-space consisting of the two spatial coordinates in the plane of the sky and the radial velocity as the only available kinematic component. Thus, signals could either be located on the same line of sight as the disc emission of M31 but at different radial velocities, or they could be located outside the disc at any radial velocity. At the same time, we did not consider objects with $v_{\rm LSR} > {-140}~\mathrm{km \, s}^{-1}$ to avoid confusion with Galactic foreground emission.
  
  It is important to note that we have to change the definition of `high-velocity cloud' for our M31 observations. The HVCs around the Milky Way were characterised by their peculiar radial velocities as seen from a point of view within the Galactic \ion{H}{i} disc. In the case of M31, however, we are looking at the entire galaxy and its HVC population from the outside. Therefore, the term `HVC' refers to all circumgalactic gas clouds which are thought to be the equivalents of the circumgalactic gas clouds around the Milky Way known as `high-velocity clouds'. This includes clouds whose radial velocities are compatible with the rotation curve of M31 but which are located well beyond the edge of the \ion{H}{i} disc.

  \begin{figure}
    \centering
    \includegraphics[width=\linewidth]{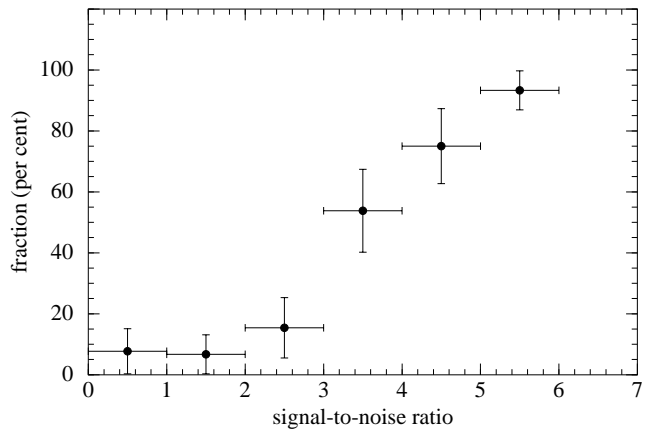}
    \caption{Fraction of detected spectral lines as a function of signal-to-noise ratio, derived from the investigation of 500~artificial spectra as described in the text. About 50~per~cent of all signals with a peak intensity of $3.5 \, \sigma$ were detected. Above $5 \, \sigma$ we are virtually complete with only a single of the 14~lines above $5 \, \sigma$ being missed.}
    \label{fig_fluxcomp}
  \end{figure}
  
  \begin{figure*}
    \includegraphics[width=0.672\linewidth]{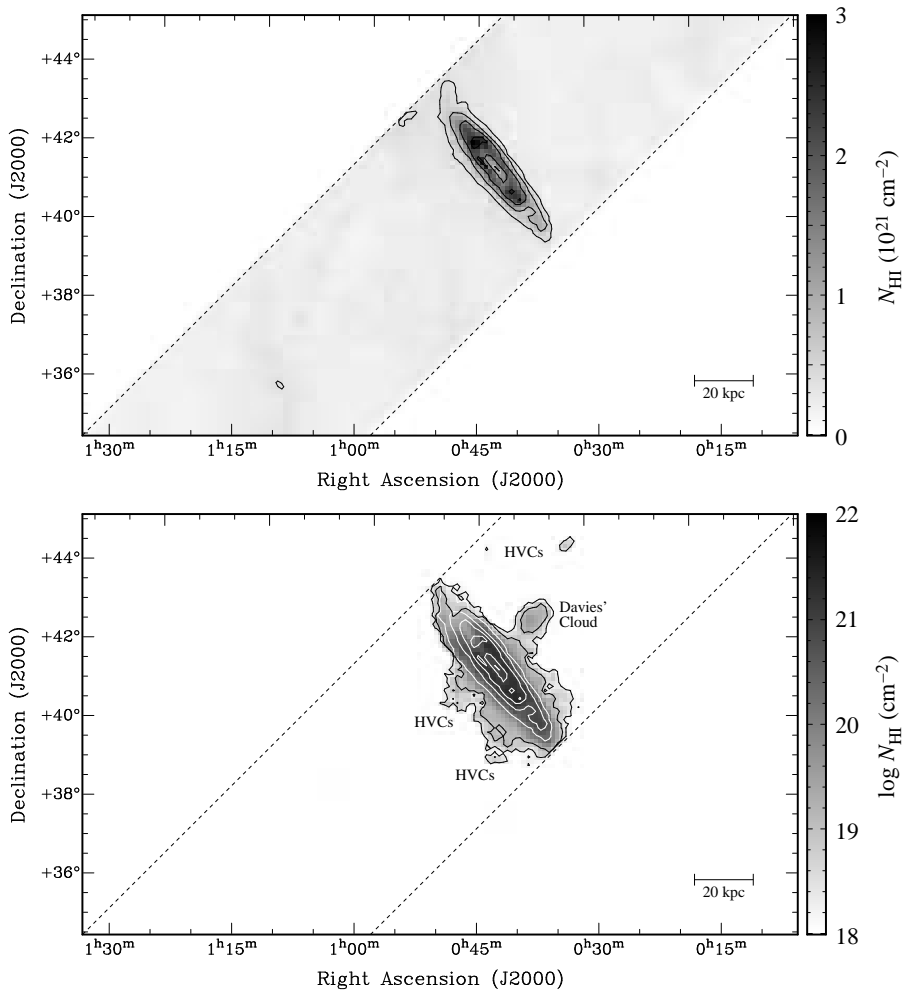}
    \caption{\ion{H}{i} column density maps of our Effelsberg blind survey of M31. \textbf{Top:} Integrated \ion{H}{i} column density in the velocity range of $v_{\rm LSR} = -620 \ldots {-24}~\mathrm{km \, s}^{-1}$. The contours were drawn at 4, 8, 16, and $25 \times 10^{20}~\mathrm{cm}^{-2}$. The faint, filamentary emission all over the map originates from Galactic \ion{H}{i}. \textbf{Bottom:} Integrated \ion{H}{i} column density in the velocity range of $v_{\rm LSR} = -620 \ldots {-139}~\mathrm{km \, s}^{-1}$, using a logarithmic intensity scale. The black contours correspond to $2 \times 10^{18}$ and $1 \times 10^{19}~\mathrm{cm}^{-2}$. The white contours are those from the upper map. Several regions of extra-planar gas and high-velocity clouds can be seen all around M31.}
    \label{fig_surveymap}
  \end{figure*}

  \section{Completeness of the data}
  \label{sect_completeness}
  
  \subsection{Spatial coverage of the map}
  
  An important question arising from the approach of a visual inspection of the survey is the question of detectability and completeness of HVC/CHVC candidates. A first limitation of the completeness of our M31 survey results from the spatial incompleteness of our map at larger radii. The geometric situation is illustrated in Fig.~\ref{fig_spatcomp}. There are five different regions for which analytic expressions for the spatial completeness have to be derived separately. The five regions are separated from each other at radii of $2\fdg{}5$, $3\fdg{}5$, $7\fdg{}9$, and $12\fdg{}7$ from the centre of M31, corresponding to projected distances of $34~\mathrm{kpc}$, $48~\mathrm{kpc}$, $108~\mathrm{kpc}$, and $174~\mathrm{kpc}$ if the distance of M31 is assumed to be $780 \; \mathrm{kpc}$ \citep{Stanek1998}. There are two ways of calculating the spatial completeness within these regions as a function of radial distance from the centre of M31. One way is to calculate the local filling factor at distance $r$ from the centre of M31 which is given by the ratio of the length of the arc intersecting the map over the total arc length, $2 \pi \, r$, of the full circle. The other way is to calculate the global filling factor of the map within a certain radius, $r$, which is the ratio of the area being covered by the map within $r$ over the total area, $\pi \, r^2$.
  
  The two spatial completeness functions, $C_{\rm L}(r)$ and $C_{\rm G}(r)$, are plotted in Fig.~\ref{fig_spatcomp2} as a function of projected distance, $r$, from the centre of M31. Within $r = 34~\mathrm{kpc}$ the survey is complete. Beyond $r = 174~\mathrm{kpc}$ the local completeness drops to $0$ whereas the global completeness decreases with $1 / r^2$. At intermediate radii both functions have a more complex behaviour. They drop abruptly at $r = 34~\mathrm{kpc}$ with the local completeness function, $C_{\rm L}(r)$, decreasing much faster than the global completeness function, $C_{\rm G}(r)$. Locally, 50~per~cent completeness is reached at a distance of $r_{50} = 48~\mathrm{kpc}$ already, whereas on the global scale the corresponding radius is $r_{50} = 77~\mathrm{kpc}$. At a projected distance from M31 of $100~\mathrm{kpc}$ the completeness is $C_{\rm L} = 12$~per~cent and $C_{\rm G} = 36$~per~cent. This means that out to $100~\mathrm{kpc}$ our survey still covers more than a third of the total area, but the local completeness at this distance is already fairly low because the survey area is concentrated around the central region.
  
  \subsection{Spatial sampling of the map}
  
  Another reason for spatial incompleteness is related to the spatial sampling of the map. The individual pointings of the map are separated by the HPBW of the telescope of $9~\mathrm{arcmin}$, so that the sensitivity is significantly decreased at positions which are located exactly in between four individual pointings. Assuming a two-dimensional Gaussian sensitivity function for the main beam, the sensitivity of each beam at the position in between four neighbouring beams (i.e.\ the corner of each grid element) is $0.25$. Averaging the four neighbouring beams results in a decrease of the RMS by a factor~2 and, thus, in a slightly higher effective sensitivity of $0.5$. As a consequence, the sensitivity for unresolved sources is in the range of~0.5 to~1, depending on the source position relative to the grid. Extended sources are covered by several pointings so that their detectability is not reduced.
  
  \begin{figure*}
    \centering
    \includegraphics[width=0.89\linewidth]{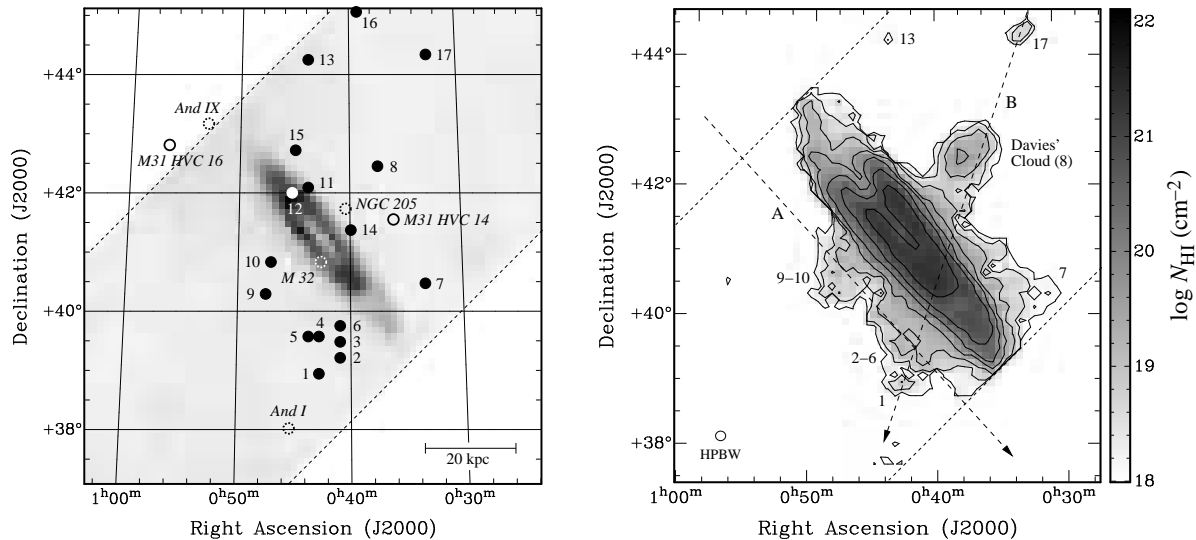}
    \caption{\textbf{Left:} Locations of the 17~individual HVCs identified in our Effelsberg survey (filled circles) superposed on an \ion{H}{i} column density map. Two HVCs studied with the WSRT \citep{Westmeier2005a} but not detected in our survey are plotted as solid open circles. The dashed open circles denote the positions of known satellite galaxies of M31. \textbf{Right:} \ion{H}{i} column density map in the velocity range of $v_{\rm LSR} = -608 \ldots {-135}~\mathrm{km \, s}^{-1}$, using the same mask as for Fig.~\ref{fig_surveymap}. The contours range from $18$ to $22~\mathrm{dex}$ in steps of $0.5~\mathrm{dex}$. The two dashed arrows indicate the lines for the position-velocity diagrams in Fig.~\ref{fig_m31effposivelo}.}
    \label{fig_m31hvcmap}
  \end{figure*}
  
  \subsection{Flux detection limit}
  \label{sec_incompflux}
  
  Another important question is down to which signal-to-noise ratio potential CHVCs can be detected. To answer this question, we simulated 500~spectra with a Gaussian RMS and a velocity resolution of $10.1 \; \mathrm{km \, s}^{-1}$. Each spectrum was multiplied with a sine function of random amplitude, wavelength, and phase shift to get arbitrary baseline shapes. With a probability of 15~per~cent a Gaussian spectral line was added to each spectrum. Line widths were randomly chosen between $15$ and $35 \; \mathrm{km \, s}^{-1}$, corresponding to what is expected for the warm neutral medium of HVCs. Peak intensities were also assigned randomly with signal-to-noise ratios between $0$ and $6$. Our algorithm added 81~artificial spectral lines with the above-mentioned parameters to the 500~spectra. Positions, intensities, and line widths of the 81~spectral lines were written to a data file for later comparison.
  
  Next, the entire data set of 500~artificial spectra was searched by eye to detect the hidden spectral lines. In total, $34$ of the $81$ signals could be identified. The resulting completeness function is shown in Fig.~\ref{fig_fluxcomp}. At a peak intensity level of $3.5 \, \sigma$ the completeness is about 50~per~cent. Above $5 \, \sigma$ we are virtually complete with only a single of the 14~lines above $5 \, \sigma$ being missed. Note that there are also two detections at about $1 \, \sigma$. In both cases strong noise peaks by chance have increased the signal above the detection threshold. These results demonstrate that, when using the classical $3 \, \sigma$ criterion for the detectability of spectral signals, we have to consider that source counts below the $5 \, \sigma$ level suffer from incompleteness. On the other hand, detections are possible even at intensities below $3 \, \sigma$ although the detection rate decreases significantly.
  
  \subsection{Confusion with Galactic emission}
  
  In optical astronomy the investigation of extragalactic objects is affected by extinction due to dust in the interstellar medium of the Galaxy. The strongest extinction is observed at low Galactic latitudes along the so-called `zone of avoidance'. The situation in \ion{H}{i} spectroscopy is slightly different. Emission lines of extragalactic objects can be superposed on those of the Milky Way, making an identification and analysis difficult or even impossible. Galactic \ion{H}{i} emission is found not only near the Galactic plane but all across the sky. However, only the radial velocity range around $v_{\mathrm{LSR}} = 0 \; \mathrm{km \, s}^{-1}$ is affected, whereas sources with radial velocities of $|v_{\mathrm{LSR}}| \gtrsim 100 \; \mathrm{km \, s}^{-1}$ are usually not subject to confusion with Galactic emission lines. In other words, the `zone of avoidance' in \ion{H}{i} spectroscopy covers the entire sky but only a limited radial velocity range for each line of sight.
  
  The systemic radial velocity of M31 is $v_{\mathrm{LSR}} \approx -300 \; \mathrm{km \, s}^{-1}$. On the approaching side of M31 we observe radial velocities down to about $-600 \; \mathrm{km \, s}^{-1}$, implying values up to about $0 \; \mathrm{km \, s}^{-1}$ on the receding side. The latter is well within the velocity range of Galactic \ion{H}{i} emission in this area which reaches out to about $-140 \; \mathrm{km \, s}^{-1}$. Hence, a major part of the receding side of M31 is blended with Galactic emission. Blending effects could also affect the detection of HVCs around M31. As the peak intensities of HVCs are expected to be small compared to the intensities of Galactic emission lines, HVCs superposed on Galactic emission are virtually undetectable. In our survey this should be the case for all potential HVCs with $v_{\mathrm{LSR}} \gtrsim -140 \; \mathrm{km \, s}^{-1}$, resulting in incompleteness with respect to certain radial velocities.

  \begin{figure}
    \centering
    \includegraphics[width=0.95\linewidth]{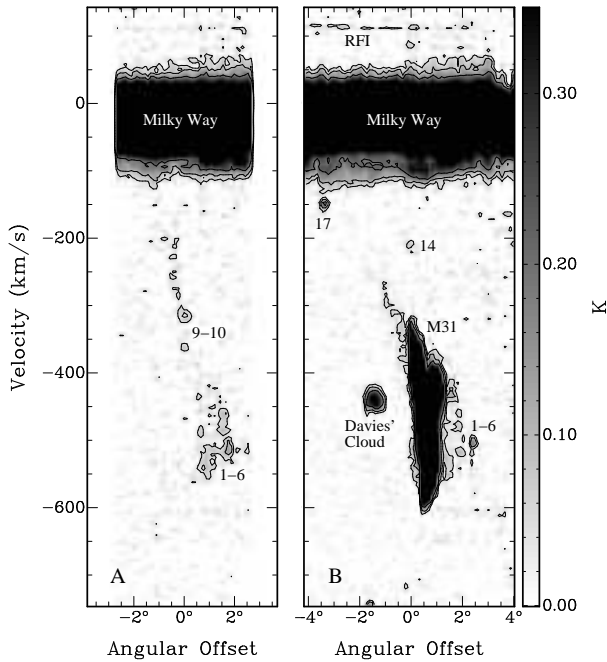}
    \caption{The two panels show position-velocity diagrams along the lines indicated by the two dashed arrows in the column density map in the right panel of Fig.~\ref{fig_m31hvcmap}. Several HVCs and regions of extra-planar gas can be seen around M31. The contour levels are $50$, $100$, and $200~\mathrm{mK}$. Note that Davies' Cloud is completely isolated in phase space although in the column density map there appears to be a connection with the \ion{H}{i} disc of M31.}
    \label{fig_m31effposivelo}
  \end{figure}

  \section{The detected population of HVCs}
  \label{sect_results}
  
\begin{figure*}
    \centering
    \includegraphics[width=0.875\linewidth]{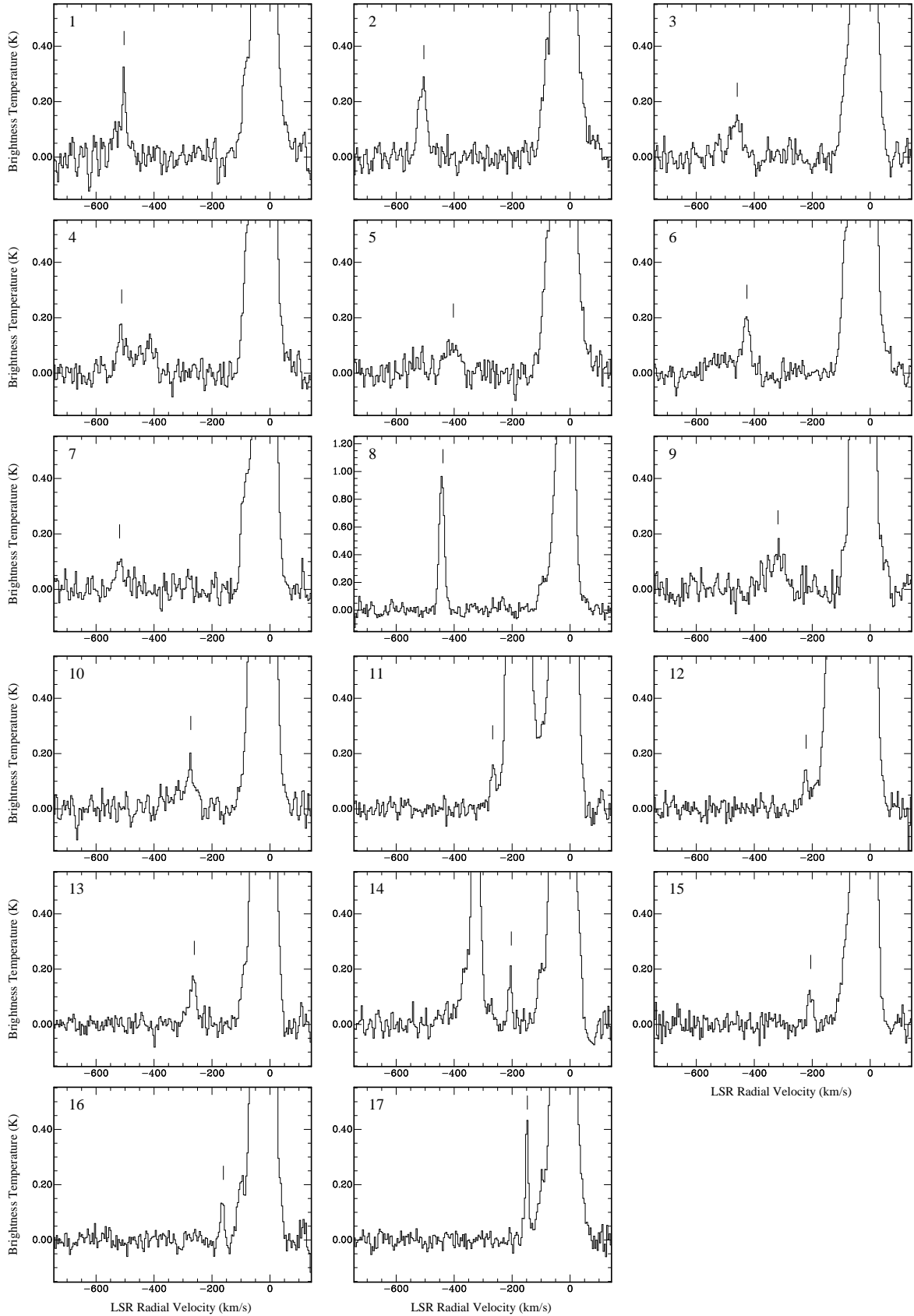}
    \caption{Effelsberg \ion{H}{i} spectra of the HVCs around M31 according to Table~\ref{tab_m31_hvcs}. The position of the spectral line belonging to the HVC is marked with a vertical line in each spectrum. The strong emission at $v_{\rm LSR} \approx 0 \; \mathrm{km \, s}^{-1}$ is usually due to \ion{H}{i} gas in the Galactic disc. In a few spectra, however, emission from the M31 disc is also present at different velocities.}
    \label{fig_effspec}
  \end{figure*}

  In total we identified 17~discrete HVCs in our Effelsberg survey. Each HVC was later confirmed by follow-up observations with the Effelsberg telescope. An overview of all clouds and their observational parameters is given in Table~\ref{tab_m31_hvcs} and Fig.~\ref{fig_surveymap} and~\ref{fig_m31hvcmap}. All HVCs are located within a projected distance of about $50~\mathrm{kpc}$ from M31. Some of them are isolated in position and velocity, whereas others appear to be associated with the diffuse extra-planar gas near the edge of the \ion{H}{i} disc of M31. A few clouds, such as number~14 in our catalogue, are seen in projection against the \ion{H}{i} disc of M31 and can only be discriminated by their different radial velocities. This situation is illustrated by the position-velocity diagrams in Fig.~\ref{fig_m31effposivelo}. \ion{H}{i} spectra of all clouds are shown in Fig.~\ref{fig_effspec}.
  
  \ion{H}{i} column density maps of the entire survey area are displayed in Fig.~\ref{fig_surveymap}. The upper map shows the entire flux integrated over a velocity range of $v_{\rm LSR} = -620 \ldots {-24} \; \mathrm{km \, s}^{-1}$. In the direction of M31 the \ion{H}{i} emission of the Milky Way disc reaches out to velocities of $v_{\rm LSR} \approx -140 \; \mathrm{km \, s}^{-1}$, resulting in a significant overlap with emission from the north-eastern part of M31. Filamentary emission from the Galactic disc can be seen all over the map with typical column densities of a few times $10^{20} \; \mathrm{cm}^{-2}$. Two conspicuous features of M31, the famous \ion{H}{i} ring and the warp of the \ion{H}{i} disc, are immediately visible in the map. The ring was first clearly detected by \citet{Roberts1966} and has a radius of approximately $10 \; \mathrm{kpc}$. The central depression in the \ion{H}{i} disc shows column densities of about 30~per~cent of the peak values found along the ring \citep{Guibert1974}. The warping of M31 can be seen towards the far ends of the disc, in particular in the north-eastern part, where the so-called `perturbed outer arm' \citep{Whitehurst1978,Sawa1982} is significantly bent in the northern direction.
  
  \begin{table*}
    \centering
    \caption{List of the 17~discrete HVCs around M31 detected in our \ion{H}{i} survey with the Effelsberg telescope. The columns denote the catalogue number of the cloud, its name according to \citet{Westmeier2005a}, right ascension, $\alpha$, and declination, $\delta$, the projected distance, $r_{\rm proj}$, of the column density maximum from the centre of M31, the LSR radial velocity, $v_{\rm LSR}$, the FWHM of the spectral line, $\Delta v$, the peak \ion{H}{i} column density, $N_{\rm H\,I}$, and the \ion{H}{i} mass, $M_{\rm H\,I}$. Radial velocities are accurate within about $\pm 5~\mathrm{km \, s}^{-1}$, and the statistical uncertainties of line widths, column densities, and \ion{H}{i} masses are of the order of 10~per~cent.}
    \label{tab_m31_hvcs}
    \begin{tabular}{rlrrrrrrr}
      \hline
        \#  &  Name  &  $\alpha$  &  $\delta$  &  $r_{\rm proj}$  &  $v_{\rm LSR}$  &  $\Delta v$  &  $N_{\rm H\,I}$  &  $M_{\rm H\,I}$ \\
            &  &  (J2000)  &  (J2000)  &  (kpc)  &  ($\mathrm{km \, s}^{-1}$)  &  ($\mathrm{km \, s}^{-1}$)  &  ($10^{19}~\mathrm{cm}^{-2}$)  &  ($10^5~{\rm M}_{\odot}$) \\
      \hline
         1  &  M31~HVC~1    &  $00^{\rm h}43^{\rm m}00^{\rm s}$  &  $38^{\circ}53'$  &  $32.5$  &  $-501$  &  $17$  &   $1.1$  &  $7.8$ \\
         2  &  M31~HVC~2    &  $00^{\rm h}41^{\rm m}06^{\rm s}$  &  $39^{\circ}16'$  &  $27.6$  &  $-509$  &  $34$  &   $1.7$  &  $6.3$ \\
         3  &  M31~HVC~7    &  $00^{\rm h}41^{\rm m}06^{\rm s}$  &  $39^{\circ}28'$  &  $24.9$  &  $-463$  &  $47$  &   $1.2$  &  $3.9$ \\
         4  &  M31~HVC~5    &  $00^{\rm h}43^{\rm m}00^{\rm s}$  &  $39^{\circ}31'$  &  $23.9$  &  $-507$  &  $34$  &   $0.9$  &  $6.3$ \\
         5  &  M31~HVC~12   &  $00^{\rm h}43^{\rm m}17^{\rm s}$  &  $39^{\circ}35'$  &  $23.0$  &  $-409$  &  $58$  &   $1.1$  &  $4.3$ \\
         6  &  M31~HVC~10   &  $00^{\rm h}41^{\rm m}05^{\rm s}$  &  $39^{\circ}41'$  &  $22.0$  &  $-426$  &  $26$  &   $0.9$  &  $4.7$ \\
         7  &  --           &  $00^{\rm h}33^{\rm m}18^{\rm s}$  &  $40^{\circ}32'$  &  $26.2$  &  $-518$  &  $39$  &   $0.7$  &  $3.1$ \\
         8  & Davies' Cloud &  $00^{\rm h}37^{\rm m}25^{\rm s}$  &  $42^{\circ}26'$  &  $20.7$  &  $-442$  &  $21$  &   $3.9$  &  $130$ \\
         9  &  --           &  $00^{\rm h}47^{\rm m}43^{\rm s}$  &  $40^{\circ}19'$  &  $18.3$  &  $-324$  &  $71$  &   $1.6$  &  $5.9$ \\
        10  &  --           &  $00^{\rm h}47^{\rm m}12^{\rm s}$  &  $40^{\circ}51'$  &  $12.9$  &  $-282$  &  $67$  &   $1.4$  &  $5.0$ \\
        11  &  --           &  $00^{\rm h}43^{\rm m}53^{\rm s}$  &  $42^{\circ}07'$  &  $11.9$  &  $-266$  &  $16$  &   $0.4$  &  $1.2$ \\
        12  &  --           &  $00^{\rm h}45^{\rm m}35^{\rm s}$  &  $42^{\circ}00'$  &  $12.3$  &  $-223$  &  $13$  &   $0.3$  &  $1.0$ \\
        13  &  M31~HVC~15   &  $00^{\rm h}43^{\rm m}55^{\rm s}$  &  $44^{\circ}13'$  &  $40.2$  &  $-266$  &  $30$  &   $0.9$  &  $3.5$ \\
        14  &  M31~HVC~13   &  $00^{\rm h}39^{\rm m}56^{\rm s}$  &  $41^{\circ}16'$  &  $ 7.1$  &  $-207$  &  $14$  &   $0.5$  &  $6.4$ \\
        15  &  --           &  $00^{\rm h}45^{\rm m}02^{\rm s}$  &  $42^{\circ}45'$  &  $21.0$  &  $-206$  &  $21$  &   $0.5$  &  $1.7$ \\
        16  &  --           &  $00^{\rm h}39^{\rm m}09^{\rm s}$  &  $45^{\circ}04'$  &  $52.5$  &  $-165$  &  $17$  &   $0.6$  &  $2.2$ \\
        17  &  --           &  $00^{\rm h}32^{\rm m}42^{\rm s}$  &  $44^{\circ}20'$  &  $48.7$  &  $-149$  &  $11$  &   $0.9$  &  $8.2$ \\
      \hline
    \end{tabular}
  \end{table*}
  
  The lower map in Fig.~\ref{fig_surveymap} is a high-contrast logarithmic map of \ion{H}{i} column densities of $10^{18} \ldots 10^{22}~\mathrm{cm}^{-2}$ in the velocity range of $v_{\rm LSR} = {-620} \ldots {-139} \; \mathrm{km \, s}^{-1}$. Here, the velocity range of Galactic \ion{H}{i} emission has been excluded with the restriction that some of the emission from the north-eastern end of the disc of M31 is also missing. In addition, we created a mask data cube which was smoothed in position and velocity. Only those mask elements with intensities in excess of $17.5~\mathrm{mK}$ were selected in the original data cube to contribute to the moment~0 map, resulting in a high-quality image showing \ion{H}{i} column densities down to a few times $10^{18}~\mathrm{cm}^{-2}$.
  
  Apart from regular disc emission, several regions of extra-planar gas and isolated \ion{H}{i} clouds are visible. Extra-planar gas emission is particularly prominent along the south-eastern edge of the disc, but extended, diffuse \ion{H}{i} emission with column densities of $N_{\rm H\,I} \lesssim 10^{19} \; \mathrm{cm}^{-2}$ is also present along the north-western edge. The most outstanding isolated \ion{H}{i} cloud in the map is Davies' Cloud located about $1\fdg{}5$ north-west of the centre of M31. The cloud was discovered by \citet{Davies1975} and studied in detail by \citet{deHeij2002a} with the WSRT. A few more compact and isolated clouds can be seen all around M31.
  
  Most of the clouds seen in Fig.~\ref{fig_surveymap} had already been discovered by \citet{Thilker2004} in their \ion{H}{i} survey of M31 with the Green Bank Telescope. A comparison between their fig.~2 and the lower map in Fig.~\ref{fig_surveymap} reveals a good correlation between the clouds seen in both surveys with the exception of a few faint and extended clouds detected in the GBT data but not seen in the Effelsberg map. Those clouds without counterparts in the Effelsberg survey are typically very diffuse with \ion{H}{i} column densities of the order of only $10^{18} \; \mathrm{cm}^{-2}$. They were only detected in the GBT data after decreasing the velocity resolution significantly to $72 \; \mathrm{km \, s}^{-1}$. With the frequency switching method used for our Effelsberg observations such diffuse objects can hardly be detected. Instead, our survey was optimised for the detection of compact sources with narrow spectral lines.
  
  Within the $3 \, \sigma$ mass detection limit of our survey of $M_{\rm H\,I} \approx 8 \times 10^4~{\rm M}_{\odot}$ we did not find any extended population of hundreds of CHVCs as observed around the Milky Way. In particular, we did not detect any HVCs or CHVCs beyond a projected distance of about $50~\mathrm{kpc}$ from M31, although our survey reaches out to significantly larger projected distances. Apparently, the detected HVCs are concentrated near M31 and less numerous than expected from the population of hundreds of HVCs/CHVCs observed around our Milky Way.
  
  \subsection{Parameters of the detected HVCs}
  
  \subsubsection{Velocities}
  
  The radial velocities of the HVCs are in the range of $v_{\rm LSR} \approx -520 \ldots \allowbreak {-150}~\mathrm{km \, s}^{-1}$. The distribution of radial velocities is not perfectly symmetric with respect to the systemic velocity of M31 of $v_{\rm LSR} \approx -300~\mathrm{km \, s}^{-1}$ (Fig.~\ref{fig_histograms}). First of all, confusion with the \ion{H}{i} emission of the Milky Way does not allow us to detect any HVCs at velocities above $-140~\mathrm{km \, s}^{-1}$, explaining the upper limit in the observed velocity range. Therefore, one or two more HVCs beyond this velocity limit could be missing in our sample due to confusion with Galactic emission. On the other hand, several of the detected clouds are arranged in a complex of HVCs near the south-eastern edge of the disc of M31 with very similar radial velocities in the range of about $-500 \ldots {-400}~\mathrm{km \, s}^{-1}$.
  
  The possibility of confusion with Galactic \ion{H}{i} emission immediately raises the question whether HVCs near Galactic velocities might actually be associated with the Milky Way rather than M31. The detection of very compact and faint gas clouds and HVCs in the halo of the Milky Way (e.g., \citealt{Lockman2002,Bruens2004,Hoffman2004,Richter2005}) has shown that the criteria of small angular size and low \ion{H}{i} column density are not sufficient to discriminate between HVCs near the Galaxy and M31. Given the lack of distance information, it is in principle impossible to associate each individual cloud with M31. The observational parameters of the detected HVCs, however, allow us to definitely associate the population as a whole with the Andromeda Galaxy. First of all, the clouds are clearly concentrated around M31 with projected distances of less than about $50~\mathrm{kpc}$. If they were associated with the Milky Way we would expect them to be more equally spread across the survey area. The probability to have them distributed in a $50~\mathrm{kpc}$ circle around M31 by chance is of the order of only $10^{-6}$ in the case of a random distribution across the sky. Second, the observed radial velocities are nearly uniformly distributed around the systemic velocity of M31 of $-300~\mathrm{km \, s}^{-1}$, although we are probably missing one or two HVCs with velocities above $-140~\mathrm{km \, s}^{-1}$. Furthermore, those HVCs with velocities of about $-500~\mathrm{km \, s}^{-1}$ would have significantly higher negative velocities than almost any other HVC near the Milky Way. Therefore, both the spatial and kinematic parameters of the detected HVCs strongly suggest that the population as a whole is associated with M31, although we cannot make this statement for individual clouds.
  
  \subsubsection{Line widths}
  
\begin{figure}
    \centering
    \includegraphics[width=\linewidth]{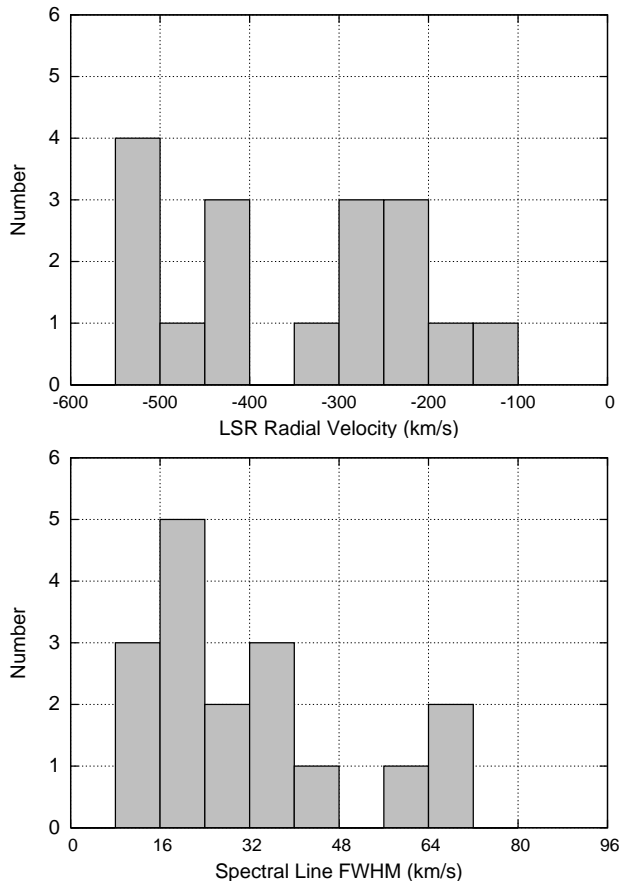}
    \caption{Histograms of radial velocities (top) and line widths (bottom) of the 17~individual HVCs identified in our survey.}
    \label{fig_histograms}
  \end{figure}

  The observed line widths of the HVCs cover a very large range from about $10$ to $70~\mathrm{km \, s}^{-1}$~FWHM (Fig.~\ref{fig_histograms}). Line widths of less than $20~\mathrm{km \, s}^{-1}$ indicate the possible presence of a cold neutral gas component or a multi-phase cold/warm neutral medium which is probably unresolved by our $9~\mathrm{arcmin}$~HPBW. A multi-phase medium is commonly found in Galactic HVCs and CHVCs (see, e.g., \citealt{Cram1976,Bruens2000,Bruens2001,Westmeier2005b}). It can also be reproduced in hydrostatic simulations of HVCs \citep{Sternberg2002} so that our evidence for the presence of a multi-phase medium in HVCs around M31 is not surprising. Difficult to explain, however, is the detection of extremely broad lines in some of the clouds. In a few cases line widths as large as $60$ or $70~\mathrm{km \, s}^{-1}$~FWHM are observed which can no longer be explained by thermal line broadening alone as the required temperatures would exceed $10^5~\mathrm{K}$. The most likely explanation for large line widths in excess of $30$ or $40~\mathrm{km \, s}^{-1}$~FWHM is a high degree of internal kinematics within the area covered by the $9~\mathrm{arcmin}$~HPBW of the Effelsberg telescope (about $2~\mathrm{kpc}$ in projected size at the distance of M31). On the one hand, several clouds or clumps with slightly different radial velocities could be aligned along the line of sight. On the other hand, individual clouds could have regular or turbulent internal motions such as global rotation or gas flows.
  
  \subsubsection{Column densities and masses}
  
  The detected \ion{H}{i} column densities are typically of the order of $10^{19}~\mathrm{cm}^{-2}$. This very low value indicates that most of the HVCs are probably not resolved by the Effelsberg telescope so that the column densities listed in Table~\ref{tab_m31_hvcs} have to be considered lower limits averaged over the $9~\mathrm{arcmin}$~HPBW of the telescope. In fact, our high-resolution synthesis observations of some of the HVCs with the WSRT \citep{Westmeier2005a} revealed significantly higher peak column densities comparable to what is observed in Galactic HVCs. Assuming a distance of $780~\mathrm{kpc}$ \citep{Stanek1998}, the resulting \ion{H}{i} masses of the HVCs are typically in the range of a few times $10^5~{\rm M}_{\odot}$ with a mean value of $4.5 \times 10^5~{\rm M}_{\odot}$. This excludes Davies' Cloud which -- if at the distance of M31 -- would have a significantly higher \ion{H}{i} mass of $1.3 \times 10^7~{\rm M}_{\odot}$. In fact, Davies' Cloud would contribute about 65~per~cent of the total mass of all detected HVCs of $2 \times 10^7~{\rm M}_{\odot}$. The total \ion{H}{i} mass of all individual HVCs and regions of extra-planar gas around M31 together is of the order of $5 \times 10^7~{\rm M}_{\odot}$. The true value will be slightly higher as some of the gas may be missing due to blending with the \ion{H}{i} emission of the Milky Way and the disc of M31 itself.
  
  It is interesting at this point to compare the \ion{H}{i} masses of the HVCs around M31 with those of Galactic HVC complexes. For several complexes, namely Complex~M \citep{Danly1993,Ryans1997}, Complex~A \citep{Wakker1996,vanWoerden1999}, Complex~WB \citep{Thom2006}, Complex~WE \citep{Wakker2001}, Complex~C \citep{Wakker2007a,Thom2008}, and the Cohen Stream and Complex~GCP \citep{Wakker2007b}, distance brackets have been published so far. These complexes are typically located a few kpc above the Galactic plane, implying \ion{H}{i} masses in the range of $\log (M_{\rm H\,I} / {\rm M}_{\odot}) \approx 4.8 \ldots 6.7$. This mass range is consistent with the \ion{H}{i} masses observed for the HVCs around M31, supporting our conclusion that they represent the analogues of the large HVC complexes seen around the Milky Way. Only the Magellanic Stream and the Leading Arm have significantly higher masses. Due to the lack of distance information, their masses can only be estimated from numerical simulations in connection with \ion{H}{i} observations. \citet{Connors2006} derived \ion{H}{i} masses of $2.4 \times 10^8~{\rm M}_{\odot}$ and $7.3 \times 10^7~{\rm M}_{\odot}$ for the Magellanic Stream and the Leading Arm, respectively, which are by orders of magnitude higher than those of the HVCs around M31, including Davies' Cloud. The Magellanic Clouds, however, are a special case with no equivalent in the M31 system.
  
  \subsubsection{Dynamical masses and external pressure}
  
\begin{figure}
    \centering
    \includegraphics[width=\linewidth]{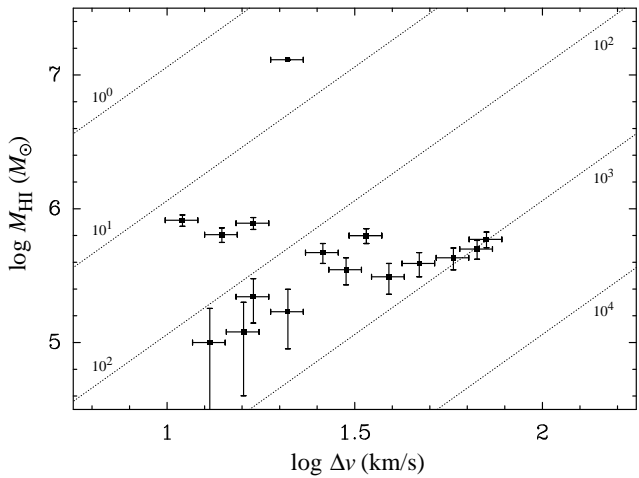}
    \caption{\ion{H}{i} mass, $M_{\rm H\,I}$, plotted against the FWHM of the spectral line, $\Delta v$, for all HVCs detected in our Effelsberg survey. The dotted lines indicate different values of $\alpha \equiv M_{\rm vir} / M_{\rm H\,I}$ under the assumption of a constant virial radius of $R_{\rm vir} = 0.5~\mathrm{kpc}$ (see Eq.~\ref{eqn_virimass2}). Hence, clouds with a small $M_{\rm vir} / M_{\rm H\,I}$ ratio are located in the upper left region of the diagram, whereas clouds with a large $M_{\rm vir} / M_{\rm H\,I}$ ratio can be found at the lower right.}
    \label{fig_virimass}
  \end{figure}

  To get an estimate of the mass fraction of neutral hydrogen in the M31 HVCs we can compare the observed \ion{H}{i} masses of the HVCs with the FWHM of the spectral lines which, in the ideal case, should provide an estimate of the virial masses of the clouds. In Fig.~\ref{fig_virimass} we have plotted the \ion{H}{i} mass, $M_{\rm H\,I}$, versus the FWHM, $\Delta v$, of all HVCs detected in our sample. The error in line width is assumed to be 10~per~cent, whereas the error bars for the mass indicate our $3 \, \sigma$ mass sensitivity of $8 \times 10^4~{\rm M}_{\odot}$. If their \ion{H}{i} masses were proportional to their virial masses, all clouds would be scattered around a straight line running from the lower left to the upper right part of the diagram. In its most simple form the virial equation reads
  \begin{equation}
    \Delta v^2 = \frac{8 \ln 2}{5} \, \frac{\mathrm{G} M_{\rm vir}}{R_{\rm vir}} \approx \frac{\mathrm{G} M_{\rm vir}}{R_{\rm vir}} \, , \label{eqn_virimass}
  \end{equation}
  assuming an isothermal sphere with constant mass density and the same mass for all particles forming the cloud. We can now replace the virial mass, $M_{\rm vir}$, in Eq.~\ref{eqn_virimass} by the \ion{H}{i} mass by introducing a factor $\alpha \equiv M_{\rm vir} / M_{\rm H\,I}$ such that
  \begin{equation}
    \Delta v^2 = \alpha \, \frac{\mathrm{G} M_{\rm H\,I}}{R_{\rm vir}} \, .       \label{eqn_virimass2}
  \end{equation}
  Lines of constant virial-to-\ion{H}{i} mass ratio, $\alpha$, according to Eq.~\ref{eqn_virimass2} are also plotted in Fig.~\ref{fig_virimass}, assuming a constant virial radius of $R_{\rm vir} = 0.5~\mathrm{kpc}$ for each cloud. This value for $R_{\rm vir}$ is suggested by our follow-up observations of several of the HVCs with the WSRT \citep{Westmeier2005a}. Obviously, most of the HVCs near M31 are spread around $\alpha \approx 100$, suggesting that they could be gravitationally dominated by other mass components such as ionised gas or dark matter. However, a few clouds show very high mass ratios of $\alpha \approx 1000$. In most of these cases we observe exceptionally broad \ion{H}{i} lines with line widths exceeding $40~\mathrm{km \, s}^{-1}$~FWHM, suggesting that the clouds cannot be considered entities in dynamical equilibrium so that the observed line widths probably do not provide any meaningful estimate of the associated total mass. On the other hand, Davies' Cloud lies slightly off the other data points, resulting from its fairly narrow spectral lines combined with a very high \ion{H}{i} mass. None the less, its true mass ratio at the distance of M31 would be $\alpha \approx 26$ because its physical diameter of about $7~\mathrm{kpc}$ would be much higher than the value of $0.5~\mathrm{kpc}$ assumed in our calculations. It is important to keep in mind, however, that for all these calculations we had to assume that the clouds are at the same distance as M31.
  
  Another possibility to stabilise the clouds in addition to their own mass is through the pressure of their ambient medium. Assuming an isothermal sphere, the virial equation for a cloud with external pressure support reads
  \begin{equation}
    4 \pi R_{\rm vir}^3 P = \frac{3 \mathrm{k} T M_{\rm vir}}{\mu} - \frac{3 \mathrm{G} M_{\rm vir}^2}{5 R_{\rm vir}} \, , \label{eqn_virialmass_pressure}
  \end{equation}
  where $P$ is the external pressure, $T$ is the gas temperature, and $\mu$ denotes the mean mass per particle \citep{Spitzer1978}. Let us assume the virial mass to be equal to the \ion{H}{i} mass and of the order of $M_{\rm vir} = 10^5~{\rm M}_{\odot}$. Then we can use Eq.~\ref{eqn_virialmass_pressure} to determine the external pressure required to stabilise the cloud without any additional mass component. Assuming a virial radius of $R_{\rm vir} = 0.5~\mathrm{kpc}$, a temperature of the warm neutral medium of $T = 10^4~\mathrm{K}$, and a mean mass equal to the mass of a hydrogen atom of $\mu = 1.674 \times 10^{-27}~\mathrm{kg}$, we obtain an external pressure of
  \begin{equation}
    P / \mathrm{k} = 84~\mathrm{K \, cm}^{-3} \, .
  \end{equation}
  Therefore, an external pressure of the order of $100~\mathrm{K \, cm}^{-3}$ is sufficient to stabilise a typical HVC around M31 in addition to its own gravitational potential. In this case no additional mass components, such as molecular gas or dark matter, would be necessary. (Note that our calculation does not even account for the expected amount of primordial helium.) The required pressure is of the order of what is expected for the ionised coronal gas around massive spiral galaxies with a temperature of about $10^6~\mathrm{K}$ and a typical density of $10^{-4}~\mathrm{cm}^{-3}$ (e.g., \citealt{Baldwin1954,Sembach2003,Rasmussen2003}). This result demonstrates that the HVCs observed around M31 can be in pressure equilibrium with the circumgalactic hot corona without the need for any additional dark matter component for stabilisation.
  
  \begin{figure*}
    \centering
    \includegraphics[width=0.82\linewidth]{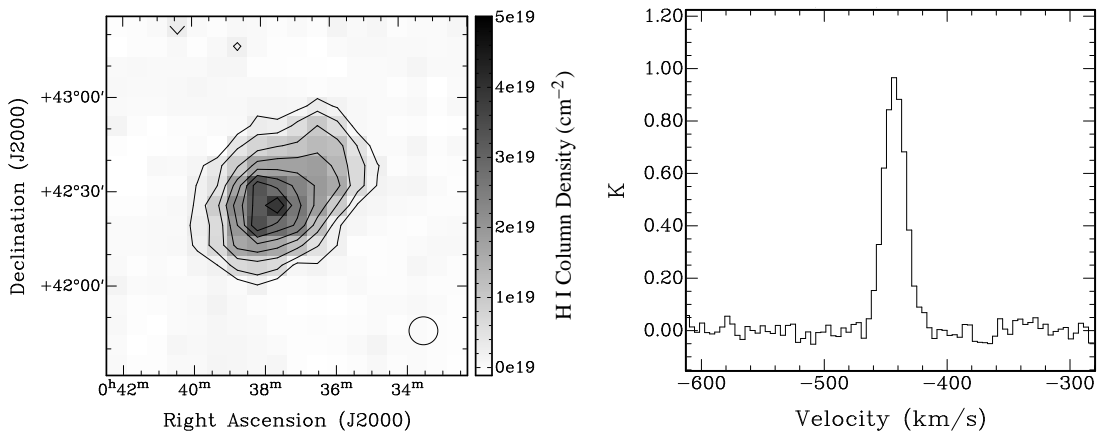}
    \caption{\ion{H}{i} column density map of Davies' Cloud. The outermost contour corresponds to $2 \times 10^{18} \; \mathrm{cm}^{-2}$, and the remaining contour lines are drawn in steps of $5 \times 10^{18} \; \mathrm{cm}^{-2}$, starting at $5 \times 10^{18} \; \mathrm{cm}^{-2}$. On the right hand side the \ion{H}{i} spectrum at the position of the column density maximum is shown. With an \ion{H}{i} mass of $1.3 \times 10^7 \; {\rm M}_{\odot}$ Davies' Cloud would be the by far most massive HVC near M31.}
    \label{fig_daviescloud}
  \end{figure*}
  
  \subsection{Discussion of individual objects}
  
  \subsubsection{Davies' Cloud}
  
  As mentioned before, Davies' Cloud would be the by far most massive individual HVC, if it were at the same distance as M31. Discovered by \citet{Davies1975}, it was studied in detail with the WSRT by \citet{deHeij2002a}. It is located only about $1\fdg{}5$ north-west of M31, corresponding to a projected distance of only $20~\mathrm{kpc}$. With an angular diameter of about $30~\mathrm{arcmin}$~FWHM, corresponding to a physical size of roughly $7~\mathrm{kpc}$ at the distance of M31, Davies' Cloud is also the most extended HVC near M31. Its appearance is not perfectly symmetric, but the column density maximum is slightly shifted in the direction of M31 (see Fig.~\ref{fig_daviescloud} for a column density map and example spectrum). The resulting head-tail morphology could be an indication for a distortion of Davies' Cloud by the ram pressure of its ambient medium. In the column density map in Fig.~\ref{fig_m31hvcmap} Davies' Cloud appears to be connected with the disc emission of M31. However, there is no such connection in phase space because the gas in the putative `bridge' has completely different radial velocities and does not provide any connection between Davies' Cloud and M31. This isolation of Davies' Cloud in phase space is obvious from the position-velocity diagram in the right panel of Fig.~\ref{fig_m31effposivelo}.
  
  Whether Davies' Cloud is indeed associated with M31 or whether it is a foreground object in the vicinity of the Milky Way has been discussed in some detail by \citet{deHeij2002a}. In their high-resolution synthesis observations with the WSRT they resolve the cold gas component of Davies' Cloud into an extended arc of compact clumps stretching mainly across the eastern edge of the cloud. This configuration, in connection with the asymmetry of the diffuse neutral gas component, suggests that Davies' Cloud is interacting with an ambient low-density medium, such as the gaseous halo or corona of M31.
  
  However, another possibility, originally discussed by \citet{Davies1975}, is an association of Davies' Cloud with the Magellanic Stream. \citet{Braun2004} detected a faint northern extension of the Magellanic Stream in their WSRT total-power survey of a large area in the region around M31 and M33. In fact, Davies' Cloud lies within only a few degrees of parts of the stream. Although typical column densities of the stream in this region are much lower than that of Davies' Cloud, the radial velocity of Davies' Cloud ($v_{\rm LSR} = {-442}~\mathrm{km \, s}^{-1}$) is very similar to the velocities of the gas associated with the Magellanic Stream ($v_{\rm LSR} \simeq {-450}~\mathrm{km \, s}^{-1}$). Therefore, a connection of Davies' Cloud with the Magellanic Stream is plausible, and Davies' Cloud may be a foreground object in the vicinity of the Milky Way. Consequently, Davies' Cloud may be less massive than previously assumed. Unfortunately, the distance of the far end of the Magellanic Stream is not well constrained, but if we consider a distance of the order $50~\mathrm{kpc}$ for Davies' Cloud,\footnote{Recent Arecibo observations by \citet{Stanimirovic2008} suggest a distance of the order of $70~\mathrm{kpc}$ for the far northern end of the Magellanic Stream.} its \ion{H}{i} mass would only be about $5 \times 10^4~{\rm M}_{\odot}$.
  
  In fact, there is more evidence for Davies' Cloud being part of the Magellanic Stream. At the distance of M31 the \ion{H}{i} mass of Davies' Cloud would be $1.3 \times 10^7~{\rm M}_{\odot}$, which is comparable to the largest HVC complexes near the Milky Way. However, while most of the HVC complexes near the Milky Way have a very elongated and filamentary structure, Davies' Cloud appears more or less spherical. This morphological difference is surprising. It is very unlikely that the cloud is significantly elongated along the line of sight as in this case we would expect to see very complex and broad spectral line profiles caused by a complex spatial and kinematic structure along the filament. Instead, the observed \ion{H}{i} lines are rather narrow with $\Delta v \approx 20~\mathrm{km \, s}^{-1}$ (FWHM) and of Gaussian shape (Fig.~\ref{fig_daviescloud}), confirming our impression of an overall spherical morphology of Davies' Cloud.
  
  \begin{figure*}
    \centering
    \includegraphics[width=0.85\linewidth]{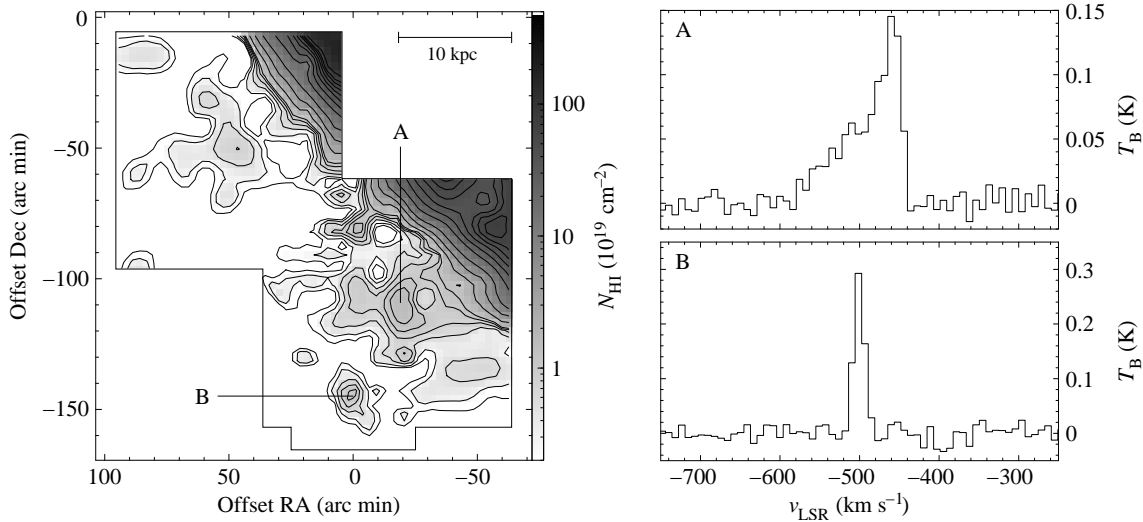}
    \caption{\ion{H}{i} column density map of the extra-planar gas found near the eastern and south-eastern edge of the disc of M31. The centre of M31 is located at offset $(0,0)$. Contours are drawn at 0.1, 0.25, 0.5, 0.75, 1, 1.25, 2.5, 5, 10, 20, 30, 40, 50, 60, 70, 80, and $100 \times 10^{19} \; \mathrm{cm}^{-2}$. Two example spectra at the positions labelled with A and B are shown on the right. The complex line profile at position~A indicates the presence of compact substructures unresolved with the $9~\mathrm{arcmin}$~HPBW of the Effelsberg telescope. The cloud at position~B shows high intensities and narrow lines of only $15 \; \mathrm{km \, s}^{-1}$ FWHM.}
    \label{fig_m31fields}
  \end{figure*}
  
  Instead, the morphology of Davies' Cloud is reminiscent of compact HVCs near the Milky Way and in particular the numerous compact clouds in the vicinity of the Magellanic Stream and the Leading Arm. \citet{BenBekhti2006} studied two of these clouds near the Leading Arm with Parkes and the ATCA. They detected numerous small clumps in the heads of the clouds similar to those found in Davies' Cloud by \citet{deHeij2002a}. In fact, the similarity between fig.~4 of \citet{BenBekhti2006} and fig.~2 of \citet{deHeij2002a} is compelling. The \ion{H}{i} lines observed in Davies' Cloud have significantly larger line widths, but this can easily be explained by a superposition of clumps along the line of sight or by slightly different physical conditions, all the more so as relatively broad \ion{H}{i} lines are commonly observed in the cores of compact HVCs (e.g., \citealt{Braun2000}).
  
  Therefore, we have to conclude that Davies' Cloud is most likely associated with the Magellanic Stream and not located at the distance of M31. This is suggested by its radial velocity, its physical properties, and its morphological structure. Irrespective of its origin, we include Davies' Cloud in our analysis of M31 HVCs for the sake of completeness, albeit emphasising its special status where indicated.
  
  \subsubsection{Extra-planar gas south-east of M31}
  
  One of the most outstanding features detected in our survey is the region of extra-planar gas near the south-eastern edge of the \ion{H}{i} disc of M31. To study this area with better sensitivity we re-observed it with a longer integration time of 5~minutes per position in in-band frequency-switching mode with the Effelsberg telescope, using the new autocorrelator AK90. The parameters are otherwise similar to those of our blind survey. The resulting \ion{H}{i} column density map of the entire region is shown in the left panel of Fig.~\ref{fig_m31fields}. The total \ion{H}{i} mass of the extra-planar gas in this area is about $2 \times 10^7~{\rm M}_{\odot}$, thus even exceeding the \ion{H}{i} mass of Davies' Cloud if it were at the distance of M31.
  
  From the map it is obvious that the extra-planar gas is not totally diffuse but instead forming more or less compact concentrations and clumps. The clumpy structure is spatially not well resolved by our $9~\mathrm{arcmin}$~HPBW. However, the individual clumps are characterised by different radial velocities which are kinematically well resolved by our spectral resolution. Two example spectra are presented in the right panel of Fig.~\ref{fig_m31fields}. In many cases the spectral profiles are rather complex and vary significantly between neighbouring pointings of the map, indicating the presence of several individual clouds or clumps with angular sizes of a few arc minutes which are unresolved by the telescope beam. We could confirm this clumpy structure in our follow-up synthesis observations of the area with the WSRT \citep{Westmeier2005a}. Those clouds which could be spatially and kinematically separated in our Effelsberg survey were included in the list of individual M31 HVCs in Table~\ref{tab_m31_hvcs}. The properties of the extra-planar gas and its possible origin in connection with the giant stellar stream \citep{Ibata2001} of M31 are discussed in detail by \citet{Westmeier2005a}.

  \section{Comparison with models and theoretical predictions}
  \label{sect_comparison}
  
  One of the most fundamental questions arising from our observations is how the detected HVC population around M31 relates to different models derived from other observations or theoretical considerations. To investigate this question we will compare the observational parameters of the HVCs in a statistical approach with different models and theoretical predictions.
  
  \subsection{The Local Group model of \citet{deHeij2002b}}
  
  Based on the LDS and HIPASS, \citet{deHeij2002b} compiled an all-sky catalogue of CHVCs. To study the distribution and origin of the CHVC population, they constructed several Local Group population models with different parameters and compared the results with the observed parameters of the CHVCs. These models were motivated by the suggestion of \citet{Blitz1999} that HVCs could be the gaseous counterparts of the `missing' dark-matter satellites predicted by CDM structure formation simulations \citep{Klypin1999,Moore1999}. The free parameters of the models by \citet{deHeij2002b} include the Gaussian dispersion, $\sigma$, of the radial distribution of CHVCs around their host galaxy in the range of $\sigma = 100~\mathrm{kpc} \ldots 2~\mathrm{Mpc}$, the slope, $\beta$, of the \ion{H}{i} mass function in the range of $\beta = {-2.0} \ldots {-1.0}$, and the highest allowed \ion{H}{i} mass, $M_{\rm max}$, of the clouds in the range of $\log \left( M_{\rm max} / {\rm M}_{\odot} \right) = 6.0 \ldots 9.0$. Using the $\chi^2$ test and the Kolmogorov--Smirnov (KS) test, \citet{deHeij2002b} determined the parameter space providing the best representation of the observational parameters of CHVCs. One of the best-fitting models is their model~\#9 with $\sigma = 200~\mathrm{kpc}$, $\beta = {-1.7}$, and $M_{\rm max} = 10^7~{\rm M}_{\odot}$.
  
  \subsubsection{Construction of a model HVC population}
  
  In order to compare our results with the parameters of model~\#9 of \citet{deHeij2002b}, we can construct a model with similar statistical parameters and then correct for the incompleteness issues of our Effelsberg observations. According to model~\#9, the total number of HVCs around M31 should be $N = 750$. This assumes equipartition of the entire model population between M31 and the Milky Way, which are both believed to have similar total mass \citep{Evans2000b}. The spatial distribution of the clouds has to be spherically-symmetric with a Gaussian radial decline. Following the method introduced by \citet{Box1958}, we can assign each cloud in our model a radial distance from M31 of
  \begin{equation}
    r = \left| \sqrt{{-2} \sigma^2 \ln (x_1)} \cos (2 \pi x_2) \right|
  \end{equation}
  with $\sigma$ being the radial dispersion of the spatial distribution of our HVC population. Furthermore, $x_1$ and $x_2$ must be two independent uniform random deviates in the range of $0 < x_i < 1$. The remaining two spatial coordinates, the angles $\varphi$ and $\vartheta$, have to be determined such that the directions of the HVCs, as seen from the centre of M31, are distributed uniformly.
  
  The next step is to assign each of the model clouds a radial velocity $v_{\rm rad}$. Again, we can use the method of \citet{Box1958} to determine radial velocities with a Gaussian distribution around the systemic velocity of M31 of $v_{\rm M31} \approx {-300}~\mathrm{km \, s}^{-1}$. Hence,
  \begin{equation}
    v_{\rm rad} = v_{\rm M31} + \sqrt{{-2} \sigma_v^2 \ln (x_1)} \cos (2 \pi x_2)
  \end{equation}
  with $\sigma_v$ being the line-of-sight velocity dispersion of the entire population of CHVCs which is expected to be governed by the details of the mass model assumed for M31. For the sake of simplicity we can use the same value as observed for the population of globular clusters around M31, namely $\sigma_v \approx 150~\mathrm{km \, s}^{-1}$ \citep{Perrett2002}.
  
  In the final step we have to assign each model cloud an \ion{H}{i} mass such that the postulated mass function is preserved. In the model of \citet{deHeij2002b} a power-law mass function with a slope of $\beta = {-1.7}$ is assumed. We can reproduce this mass function via
  \begin{equation}
    M_{\rm H\,I} = M_{\rm min} x^{1 / (\beta + 1)} ,
  \end{equation}
  where $M_{\min}$ is the lower end of the mass function, and $x$ must again be a uniform random deviate in the range of $0 < x < 1$.
  
  \begin{figure*}
    \centering
    \includegraphics[width=0.85\linewidth]{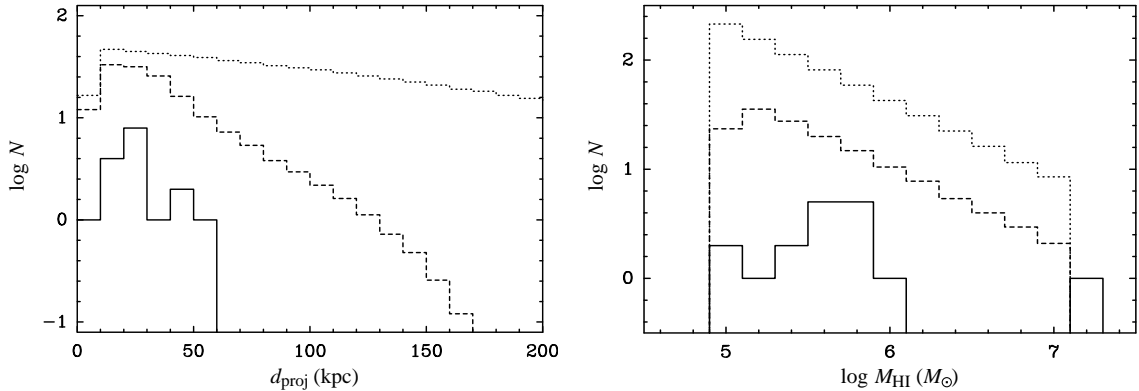}
    \caption{Projected radial distribution (left) and \ion{H}{i} mass spectrum (right) of HVCs around M31 according to our Effelsberg observations (solid line) and our model (dotted/dashed lines) based on model~\#9 of \citet{deHeij2002b}. In both diagrams the dotted line corresponds to the original model, whereas the dashed line represents the remaining model population after correcting for the different incompleteness issues of our survey.}
    \label{fig_mass_and_radius}
  \end{figure*}
  
  \subsubsection{Consideration of the survey limitation}
  
  With the ingredients listed above we are able to construct a model population of HVCs around M31 with parameters similar to model~\#9 of \citet{deHeij2002b}. Before we can compare this model population with our Effelsberg observations we have to account for the different observational limitations of our survey. The most obvious limitation is the survey area for which we can account by simply removing all clouds from the model which, when projected on the sky, are located outside the survey area. Next, we have to remove all clouds which suffer from blending with \ion{H}{i} emission of the Galactic disc, namely all clouds with radial velocities of $v_{\rm rad} > {-140}~\mathrm{km \, s}^{-1}$.
  
  A correct treatment of the flux detection limit of our survey is more complicated. To assess whether a cloud of certain mass can be detected with our survey we have to find a conversion between the \ion{H}{i} mass and the observed flux of a gas cloud. According to \citet{Schneider1997}, the total observed flux, $F$, of an unresolved gas cloud of \ion{H}{i} mass $M_{\rm H\,I}$ and distance $d$ is given by
  \begin{equation}
    \frac{F}{\mathrm{mJy \, km \, s}^{-1}} = 4.24 \times 10^3 \, \frac{M_{\rm H\,I}}{{\rm M}_{\odot}} \left( \frac{d}{\mathrm{kpc}} \right)^{-2}.
    \label{eqn_fluxmass}
  \end{equation}
   The Effelsberg spectra, however, are given in terms of brightness temperature. Flux density, $S$, and brightness temperature, $T_{\rm B}$, can be converted via 
  \begin{equation}
    T_{\rm B} = \frac{\lambda^2 S}{2 \, \mathrm{k} \, \Omega} \label{eqn_fluxconversion}
  \end{equation}
  with $\mathrm{k}$ being the Boltzmann constant and $\Omega$ the solid angle of the telescope beam. For a constant wavelength of $\lambda = 21.1 \; \mathrm{cm}$ and a circular telescope beam of HPBW $\Theta$, Eq.~\ref{eqn_fluxconversion} can be simplified to 
  \begin{equation}
    \frac{T_{\rm B}}{\mathrm{K}} = 606 \, \frac{S}{\mathrm{mJy \, beam}^{-1}} \left( \frac{\Theta}{\mathrm{arcsec}} \right)^{-2} . \label{eqn_fluxconversion2}
  \end{equation}
  Assuming the source being unresolved, we can combine Eq.~\ref{eqn_fluxmass} and \ref{eqn_fluxconversion2} to obtain a relation between the \ion{H}{i} column density and the \ion{H}{i} mass of a cloud, namely 
  \begin{equation}
    \frac{N_{\rm H\,I}}{\mathrm{cm}^{-2}} = 4.68 \times 10^{24} \, \frac{M_{\rm H\,I}}{{\rm M}_{\odot}} \left( \frac{\Theta}{\mathrm{arcsec}} \; \frac{d}{\mathrm{kpc}} \right)^{-2} , \label{eqn_coldenmascon}
  \end{equation}
  where we also made use of the conversion between the integrated brightness temperature, $\int T_{\rm B} \, \mathrm{d}v$, and the \ion{H}{i} column density, $N_{\rm H\,I}$, of a cloud under the fundamental assumption that the gas is optically thin. Assuming a Gaussian line profile with dispersion $\sigma$, we can derive the \ion{H}{i} column density sensitivity of our Effelsberg survey from the brightness temperature sensitivity via 
  \begin{eqnarray}
    N_{\rm H\,I} & = & c_1 T_{\rm B} \int \limits_{-\infty~}^{\infty} \exp \! \left( -\frac{[v - v_0]^2}{2 \, \sigma^2} \right) \mathrm{d}v \notag \\
                 & = & c_1 \sqrt{\frac{-\pi}{4 \ln 0.5}} \, T_{\rm B} \Delta v \; \approx \; c_1 T_{\rm B} \Delta v \, , \label{eqn_coldensen}
  \end{eqnarray}
  where $T_{\rm B}$ denotes the peak brightness temperature and $v_0$ the radial velocity of the spectral line. $\Delta v \approx 2.355 \, \sigma$ is the FWHM of the spectral line. If $N_{\rm H\,I}$ is given in $\mathrm{cm}^{-2}$ and $T_{\rm B} \Delta v$ in $\mathrm{K \, km \, s}^{-1}$, the corresponding conversion factor is $c_1 = 1.823 \times 10^{18}$, again assuming that the gas is optically thin.
  
  \begin{figure*}
    \centering
    \includegraphics[width=0.85\linewidth]{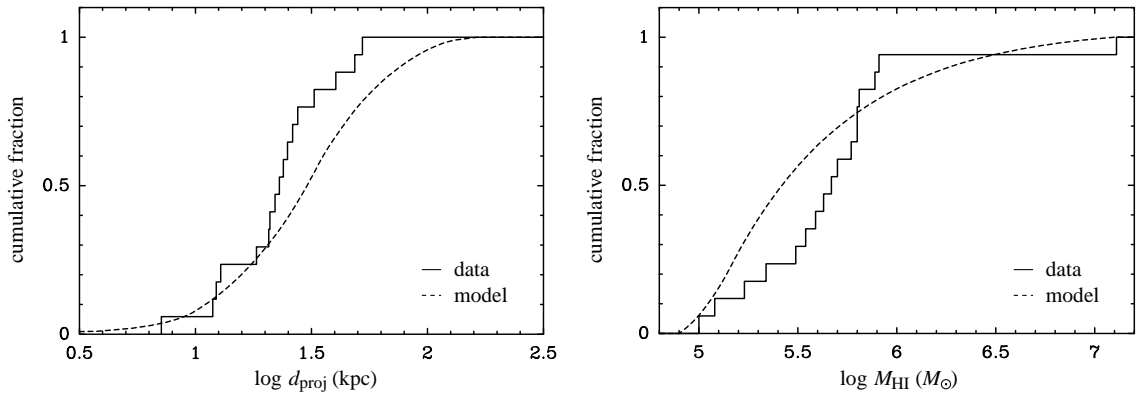}
    \caption{Normalised cumulative distribution of HVCs as a function of projected distance, $d_{\rm proj}$, from the centre of M31 (left) and \ion{H}{i} mass, $M_{\rm H\,I}$ (right). The solid line shows the 17~detected HVCs from our survey, whereas the dashed curve is the expected distribution according to our model based on model~\#9 of \citet{deHeij2002b}.}
    \label{fig_kstest}
  \end{figure*}
  
  As we have seen in Section~\ref{sec_incompflux}, the detection rate of clouds at the $3.5 \, \sigma$ level is about 50~per~cent. Below $2 \, \sigma$, the detection rate drops to zero, whereas above $5 \, \sigma$ we are virtually complete (see Fig.~\ref{fig_fluxcomp}). Thus, it is reasonable to construct a soft cut-off level with a linear increase in detection probability $P(T_{\rm B})$ such that 
  \begin{equation}
    P(T_{\rm B}) = 
    \begin{cases}
      0 & \text{for } T_{\rm B} < 2 \, \sigma \\
      \frac{T_{\rm B}}{3 \sigma} - \frac{2}{3} & \text{for } 2 \, \sigma < T_{\rm B} < 5 \, \sigma \\
      1 & \text{for } T_{\rm B} > 5 \, \sigma
    \end{cases}
    \label{eqn_detpro}
  \end{equation}
  By converting the values for $T_{\rm B}$ into the corresponding \ion{H}{i} masses (using Eq.~\ref{eqn_coldenmascon} and~\ref{eqn_coldensen}) we are able to exclude clouds from the model with a certain probability expressed by Eq.~\ref{eqn_detpro} to account for the detection limit of our Effelsberg data.
  
  \subsubsection{Comparison between model and data}
  
  To minimize statistical errors we calculated the mean parameters of 1000 model HVC populations with the specifications discussed above. Each model originally contained 750~HVCs of which on average 595 were excluded from the model population by the incompleteness issues of our Effelsberg survey, leaving us with about 155~expected HVC detections within our survey area. This number immediately illustrates that model~\#9 of \citet{deHeij2002b} overestimates the expected number of HVCs detected in our survey by about a factor~10.
  
  The resulting distribution of HVCs as a function of projected distance, $d_{\rm proj}$, from the centre of M31 is plotted in the left diagram of Fig.~\ref{fig_mass_and_radius}. The dotted curve represents the original model containing 750~HVCs. The dashed curve shows the final distribution after accounting for the incompleteness effects. The distribution drops abruptly beyond $d_{\rm proj} \approx 150~\mathrm{kpc}$ as a result of the limited survey region. The observed radial distribution is plotted for comparison as the solid line. It is obvious that both the total number and the radial distribution of the observed HVCs around M31 deviate significantly from the model. The observed HVCs are much closer to M31 than expected from the model, whereas no HVCs are observed at larger projected distances where the model still predicts several detections. To quantify these differences in radial distribution we carried out the Kolmogorov--Smirnov (KS) test on the normalised cumulative radial distribution of the modelled and observed HVCs as plotted in the left panel of Fig.~\ref{fig_kstest}. The resulting KS statistic for the two distributions is $D = 0.32$, hence rejecting the hypothesis that both samples are drawn from the same radial distribution on the 95~per~cent confidence level.
  
  The deviations are similar when considering the \ion{H}{i} mass spectrum. The modelled mass spectrum is shown in the right panel of Fig.~\ref{fig_mass_and_radius}. Again, the dotted curve is the original mass spectrum, whereas the dashed curve has been corrected for observational constraints. The modelled distribution essentially follows the power law defined by \citet{deHeij2002b}. In contrast, the observed \ion{H}{i} mass spectrum (solid line) does not appear to follow a power law at all. Instead, it peaks at $M_{\rm H\,I} \approx 5 \times 10^5~{\rm M}_{\odot}$, indicating a lack of low-mass clouds compared to the model. To compare the observed and modelled mass spectra we again used the KS test on the normalised cumulative mass distributions as plotted in the right panel of Fig.~\ref{fig_kstest}. Again, the resulting KS statistic is $D = 0.32$, thus rejecting the modelled mass function on the 95~per~cent confidence level.
  
  Obviously, the model proposed by \citet{deHeij2002b} does not provide a good and accurate description of our data. The model was motivated phenomenologically based on the population of CHVCs observed all over the sky. However, both the total number and the radial distribution of the HVCs detected around M31 do not agree with the predictions of the model. This result confirms our conclusion that HVCs/CHVCs are much closer to their host galaxies than previously assumed, implying not only smaller projected distances but also lower \ion{H}{i} masses. Hence, a large fraction of the HVC population around M31 may have \ion{H}{i} masses below our $3 \, \sigma$ detection limit of $8 \times 10^4~{\rm M}_{\odot}$ and could have remained undetected.
  
  \begin{figure*}
    \centering
    \includegraphics[width=0.85\linewidth]{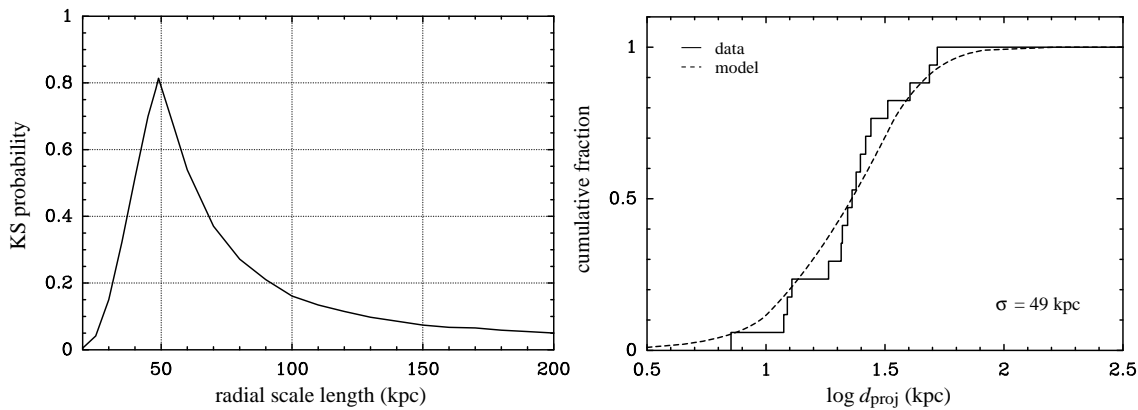}
    \caption{\textbf{Left:} Kolmogorov--Smirnov probability for the hypothesis that the observed radial distribution of HVCs around M31 follows a Gaussian distribution with radial scale length $\sigma$. There is a clear peak at $\sigma = 49~\mathrm{kpc}$ with a probability in excess of 80~per~cent. \textbf{Right:} Comparison of the normalised cumulative distributions of observed (solid line) and modelled (dashed line) HVCs as a function of projected distance from the centre of M31 for the best-fitting radial scale length of $\sigma = 49~\mathrm{kpc}$. The KS test in this case yields a statistic of $D = 0.15$, resulting in a probability of $p = 81$~per~cent that both samples were drawn from the same distribution.}
    \label{fig_bestfits}
  \end{figure*}
  
  %\begin{figure*}
  %  \centering
  %  \includegraphics[width=0.8\linewidth]{westmeier8.eps}
  %  \caption{Same as in Fig.~\ref{fig_bestfits}, this time, however, counting the HVC complex near the south-eastern edge of M31 as a single object. The radial scale length of the best-fitting model is slightly smaller, $\sigma \approx 41~\mathrm{kpc}$, with a significantly increased probability of nearly~99~per~cent.}
  %  \label{fig_bestfits2}
  %\end{figure*}
  
  \subsubsection{Modification of the original model}
  
  To study the radial distribution of the HVCs around M31 in more detail, we modified the model of \citet{deHeij2002b} in order to determine the radial scale length, $\sigma$, of the modelled HVC population which provides the best fit to the observed distribution of projected radii. In practice, we calculated several models covering the entire range of $\sigma = 20 \ldots 200~\mathrm{kpc}$, but leaving all other parameters (total number of clouds, \ion{H}{i} mass spectrum, etc.) fixed. Again, we calculated 1000~individual model populations for each value of $\sigma$ to keep statistical errors small. We then compared the projected radial distribution of the model clouds with the observed one, again using the KS test. The results are presented in Fig.~\ref{fig_bestfits}.
  
  The left diagram of Fig.~\ref{fig_bestfits} shows the KS probability for the hypothesis that the projected radial distributions of observed and modelled HVCs agree as a function of radial scale length, $\sigma$, of the model population. The probability curve shows a very pronounced and narrow peak at $\sigma \approx 49~\mathrm{kpc}$ where the probability exceeds 80~per~cent. Towards larger, and in particular towards smaller, scale lengths the KS probability drops very quickly. For $\sigma = 100~\mathrm{kpc}$ the probability is of the order of 15~per~cent, and it decreases to only about 5~per~cent for the original \citet{deHeij2002b} model of $\sigma = 200~\mathrm{kpc}$. The normalised cumulative radial distributions of model and data for the best-fitting case, $\sigma = 49~\mathrm{kpc}$, are plotted in the right panel of Fig.~\ref{fig_bestfits}. In comparison to the original model of \citet{deHeij2002b} in Fig.~\ref{fig_kstest} the radial distribution is described much better, although model and data do not match in every detail. The observed HVC population shows a significantly steeper rise in the radial profile than the model population. This steep rise is caused by the fact that 6~HVCs belong to the complex of clouds near the south-eastern edge of M31, resulting in similar projected radii. The appearance of this complex suggests that its constituents have a common origin and should not be treated as separate objects.
  
  %If we count the entire complex as one object only (Fig.~\ref{fig_bestfits2}), the best-fitting model is slightly shifted to $\sigma \approx 41~\mathrm{kpc}$ with a significantly increased probability of $p = 98.8$~per~cent indicating that we obtain a much better fit when combining all HVCs of the complex. At the same time, the peak in the probability function is broader than before, resulting in a wider range of feasible radial scale lengths. For example, at $\sigma = 100~\mathrm{kpc}$ the KS test still yields a probability of more than 25~per~cent for an agreement in the radial profiles of observed and modelled HVCs. Even for the original model of \citet{deHeij2002b} with $\sigma = 200~\mathrm{kpc}$ we still obtain a probability of more than 10~per~cent, thus excluding this model at not even 90~per~cent confidence level. When interpreting these numbers we have to keep in mind, though, that after combining all HVCs in the complex we have only a very small number of 12~HVCs remaining for our statistical analysis.
  
  In summary, the original Local Group population model of \citet{deHeij2002b} can be excluded with high confidence. Instead, our results support a more compact population with a Gaussian radial scale length of $\sigma \simeq 50~\mathrm{kpc}$. However, the radial scale length is not strictly confined by our observations. Even for $\sigma = 100~\mathrm{kpc}$ we still obtain KS probabilities of about 15~per~cent for agreement of our observations with the model.
  
  \subsection{Comparison with the satellite galaxies of M31}
  
  Like the Milky Way, M31 hosts an extended group of satellite galaxies of which M33 is the most massive one. An overview of our current census of the M31 group is given by \citet{McConnachie2006} and \citet{Koch2006}. In addition, several very faint dwarf galaxies, named Andromeda IX--XIII, have recently been discovered by \citet{Zucker2004,Zucker2006a} and \citet{Martin2006}. These detections have shown that a large number of low-mass satellites with very low surface brightness may still be awaiting discovery and that the M31 group, as presently known, is presumably incomplete.
  
  The radial distribution of satellite galaxies around M31 has been reasonably well determined (see, e.g., \citealt{Koch2006} for an overview), although the relative errors of galactocentric distances can be significant for the innermost satellites of M31. Due to the lack of distance information, the spatial distribution of the HVCs around M31, however, is completely unknown. Hence, the only feasible approach is to compare the projected positions of HVCs and satellite galaxies by discarding the distance information available for the M31 satellites.  A similar problem arises when trying to compare the mass spectrum of both populations, as for most of the M31 satellite galaxies dynamical studies are not available. There have been \ion{H}{i} studies, however, for several galaxies, although in four cases only upper limits resulting from non-detections are available (see \citealt{Westmeier2007} for a complete list of \ion{H}{i} observations and references).
  
\begin{figure}
    \centering
    \includegraphics[width=0.95\linewidth]{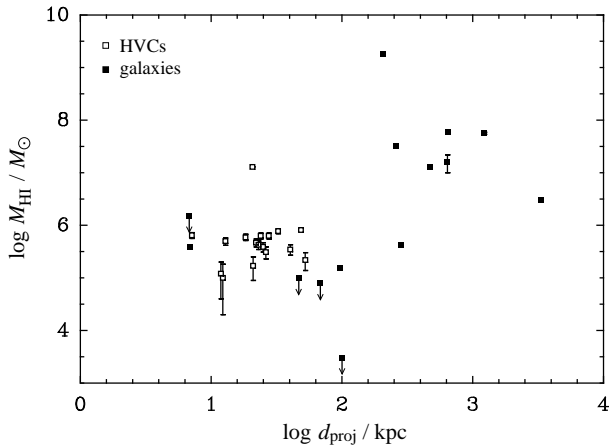}
    \caption{\ion{H}{i} mass versus projected distance of HVCs (open squares) and satellite galaxies (filled squares) around M31. Although both groups tend to populate different regions of the diagram, it is important to note that for only a fraction of the M31 satellites \ion{H}{i} studies are available.}
    \label{fig_mass_vs_dist_eff}
  \end{figure}

  In Fig.~\ref{fig_mass_vs_dist_eff} the logarithm of the \ion{H}{i} mass, $M_{\rm H\,I}$, of HVCs (open squares) and satellite galaxies (filled squares) is plotted versus the logarithm of the projected distance, $d_{\rm proj}$. Only those satellite galaxies within a projected radius of $300~\mathrm{kpc}$ around M31 have been considered as they are approximately located within the expected virial radius of M31 (see, e.g., \citealt{Kravtsov2004}). There appears to be a distinction between HVCs and satellite galaxies. The HVCs are concentrated in the lower left area of the diagram with lower \ion{H}{i} masses and smaller projected distances, whereas the galaxies tend to concentrate in the upper right region of the diagram with typically higher \ion{H}{i} masses and larger distances. Apparently, HVCs and satellite galaxies have a completely different radial distribution. All HVCs are found within a projected radius of about $50~\mathrm{kpc}$, whereas the satellite galaxies of M31 are spread across the entire range of distances out to $300~\mathrm{kpc}$. The Kolmogorov--Smirnov test yields a statistic of $D = 0.78$, resulting in a very low probability of only $1.6 \times 10^{-5}$ that HVCs and satellite galaxies are drawn from the same projected radial distribution.
  
  When interpreting the mass-distance diagram we have to keep in mind that the incompleteness and selection effects discussed above apply. As a consequence, the separation between HVCs and galaxies in Fig.~\ref{fig_mass_vs_dist_eff} may be an artefact of incomplete data. For several satellites of the M31 group, \ion{H}{i} measurements are either not available or only upper mass limits have been derived from non-detections. As this will particularly affect those galaxies with intrinsically low \ion{H}{i} masses, a large number of data points could be missing in the low-mass domain, and there could be a more significant overlap between galaxies and HVCs than suggested by Fig.~\ref{fig_mass_vs_dist_eff}.
  
  If we assume the difference between HVCs and galaxies in Fig.~\ref{fig_mass_vs_dist_eff} to be real, what could the physical reason for the observed discrepancies be? If both HVCs and satellite galaxies were the visible counterparts of dark matter satellite haloes, the differences in their radial distribution would be surprising. In this case, a physical mechanism would be required to prevent the presence of dark-matter-dominated neutral gas clouds at large galactocentric distances from M31, e.g.\ through ionisation of the gas. This scenario will be discussed in more detail in the following Sect.~\ref{sect_cdmcomp}. Another reason for the discrepancies in the radial distribution of HVCs and satellite galaxies could be a different origin of both groups of objects. If HVCs were primarily of tidal origin or the result of condensing halo gas, galactic winds, or outflows, their proximity to M31 would be a natural consequence.
  
  \subsection{Comparison with CDM simulations}
  \label{sect_cdmcomp}
  
  \citet{Kravtsov2004} carried out structure formation simulations of Galaxy-sized dark-matter haloes in the framework of $\Lambda$CDM cosmology. They also included a simple model of star formation and gas dynamics in their simulations. A comparison with their results reveals significant differences in the radial distribution of both HVCs and satellite galaxies compared to the predicted population of dark matter satellites around Galaxy-sized haloes. Within a radius of $200~h^{-1}~\mathrm{kpc}$ the mean distance of dark matter satellites of about $120~h^{-1}~\mathrm{kpc}$ is noticeably larger than that of the M31 satellite galaxies of $85~h^{-1}~\mathrm{kpc}$ and significantly larger (assuming $h \approx 0.7$) than the maximum projected distances of about $50~\mathrm{kpc}$ observed for HVCs. \citet{Kravtsov2004} can partly solve this problem by including a simple model of star formation in dark matter haloes in their simulations. According to their results only a small fraction of satellites have actually formed stars to become luminous galaxies, whereas most haloes remained dark. Their model can not only reproduce the radial distribution and number of satellite galaxies observed around the Milky Way, but also the circular velocity function and the observed morphological segregation.
  
  To investigate whether the remaining dark satellites in their simulations can be identified with the population of HVCs around M31 or the Milky Way, \citet{Kravtsov2004} also determined the gas mass associated with each dark matter halo in their simulation. Within a radius of $200~h^{-1}~\mathrm{kpc}$ they find a total gas mass of $M_{\rm g} \approx 2 \times 10^9~{\rm M}_{\odot}$, resulting in a corresponding neutral gas mass of $M_{\rm H\,I} \approx 2 \times 10^8~{\rm M}_{\odot}$ if a neutral gas fraction of 10~per~cent is assumed (e.g., \citealt{Maloney2003}). This value is one order of magnitude higher than the total \ion{H}{i} mass of about $2 \times 10^7~{\rm M}_{\odot}$ of the 17~HVCs identified in our Effelsberg survey, including Davies' Cloud. We should note, however, that some more neutral gas is present in the form of diffuse, extra-planar gas which is not resolved into individual clouds with our $9~\mathrm{arcmin}$~HPBW. %On the other hand, the observed neutral gas mass in HVCs is totally dominated by a single object, Davies' Cloud, which contributes 65~per~cent of the total \ion{H}{i} mass, assuming it is at the distance of M31.
  
  The total number of gaseous dark haloes with $M_{\rm g} > 10^6~{\rm M}_{\odot}$ (corresponding to $M_{\rm H\,I} \gtrsim 10^5~{\rm M}_{\odot}$ if a neutral hydrogen fraction of $f = 0.1$ is assumed) in the simulations of \citet{Kravtsov2004} is about $50 \ldots 100$, but for the central $50~\mathrm{kpc}$ their models predict only $2 \ldots 5$ such clouds. They conclude that part of the HVCs found around M31 by \citet{Thilker2004} could be primordial dark matter haloes, whereas others could instead be of tidal origin similar to the Magellanic Stream around the Milky Way. In our Effelsberg survey, however, we have not detected any additional HVCs beyond a projected radius of about $50~\mathrm{kpc}$ with a $3 \, \sigma$ detection limit of $8 \times 10^4~{\rm M}_{\odot}$. This discrepancy is surprising, given that more than 90~per~cent of the predicted gaseous haloes should be located outside a radius of $50~\mathrm{kpc}$.
  
  Because of the limited azimuthal range of our survey region we cannot assume to have detected all of the expected gaseous haloes. At a projected distance of $50~\mathrm{kpc}$ our local completeness function drops below 50~per~cent (see Fig.~\ref{fig_spatcomp2}), but none the less we still cover more than one third of the total area within a projected radius of $100~\mathrm{kpc}$. Within a radius of $100~\mathrm{kpc}$ \citet{Kravtsov2004} expect about 35~per~cent of all dark matter haloes, resulting in $18 \ldots 35$ expected gaseous haloes with $M_{\rm g} > 10^6~{\rm M}_{\odot}$, about a third of which should be located within our survey area, resulting in a total number of expected HVCs in our survey in the range of about $6 \ldots 12$ (not taking into account the details of the radial distribution of dark matter haloes). This is less than the 17~HVCs identified in our data, supporting the concept of \citet{Kravtsov2004} that some of the detected HVCs may represent tidally stripped gas originating from satellite galaxies of M31. However, we would expect the gaseous primordial dark matter haloes to be spread across the entire survey area, whereas the detected HVCs are highly concentrated in the central region around M31.
  
  One possible solution to this problem could be ionisation. With their masses and sizes, the HVCs observed around M31 are similar to the HVCs considered by \citet{Sternberg2002} in their hydrostatic simulations of compact HVCs. For their circumgalactic model, \citet{Sternberg2002} assumed spherical, pressure-confined, dark-matter-dominated clouds at typical distances from the Galaxy of the order of $150~\mathrm{kpc}$. The original aim of the simulations was to explain the population of CHVCs as defined by \citet{Braun1999} and to understand their observational parameters as determined by \citet{Braun2000} and \citet{Burton2001}. For the circumgalactic case, \citet{Sternberg2002} predict typical \ion{H}{i} masses of $3 \times 10^5~{\rm M}_{\odot}$ and a radial scale length of the \ion{H}{i} column density distribution of the clouds of $0.5~\mathrm{kpc}$. The typical peak \ion{H}{i} column density of their model clouds is $5 \times 10^{19}~\mathrm{cm}^{-2}$. These parameters are in very good agreement with those of the HVCs around M31 as observed with the Effelsberg telescope and the WSRT \citep{Westmeier2005a}.
  
  The HVCs modelled by \citet{Sternberg2002} are gravitationally dominated by a dark matter halo for which a Burkert profile \citep{Burkert1995} was assumed. For the circumgalactic case, \citet{Sternberg2002} assumed a virial mass of the halo of the order of $10^8~{\rm M}_{\odot}$. This implies a virial-to-\ion{H}{i} mass ratio of the order of $\alpha = 300$, which corresponds to a very small neutral gas fraction of only 0.3~per~cent. Despite the substantial dark matter content, the halo would not be massive enough to retain the warm gas. Therefore, the clouds have to be additionally stabilised by the pressure of an external medium of at least $P / \mathrm{k} \approx 50~\mathrm{K \, cm}^{-3}$. The expected circumgalactic hot ionised corona around the Milky Way and M31 could provide the required external pressure stabilisation. In this case, the central gas density of the modelled HVCs would be $N_{\rm H} \approx 2 \times 10^{-2}~\mathrm{cm}^{-3}$ which is again in excellent agreement with the observed mean neutral gas densities of the HVCs around M31 \citep{Westmeier2005a}.
  
  This remarkable agreement between the hydrostatic simulations of dark-matter-domi\-nated HVCs by \citet{Sternberg2002} and the observed parameters of the HVCs detected in our WSRT observations suggests that some of the clouds could indeed be primordial, dark-matter-dominated clouds as originally assumed by \citet{Blitz1999} and \citet{Braun1999}. This could particularly be true for those clouds at larger projected distances from M31 which are completely isolated from all other HVCs and from M31 and any of its known satellite galaxies. The required pressure confinement by the corona of M31 could also explain the head-tail morphologies observed for some of these clouds by \citet{Westmeier2005a}. As the clouds are moving with high velocities through their environment their outer parts will be distorted and stripped by the ram-pressure of the ambient medium (for hydrodynamic simulations see, e.g., \citealt{Quilis2001,Konz2002}).
  
  The simulations of \citet{Sternberg2002} now allow us to find a plausible explanation for the concentration of HVCs close to M31 in contrast to the simulations of \citet{Kravtsov2004}. At larger distances from M31 the density of the circumgalactic environment is expected to decrease. At some distance, the density could drop below the value of $P/\mathrm{k} \approx 50~\mathrm{K \, cm}^{-3}$ which is required in addition to the gravitational potential to retain the warm gas of the HVCs. This would have two consequences: First, the clouds would not be stable any more because the gas would no longer be bound. Second, the gas density would become too low to provide sufficient shielding against the ionising intergalactic radiation field. Consequently, the gas would be mainly ionised and no longer detectable in the 21-cm emission of neutral hydrogen. If this scenario is correct, hundreds of mainly ionised or pure dark-matter satellites should be present around large spiral galaxies, such as the Milky Way and M31, which would not be detectable in the 21-cm line of neutral hydrogen.
  
  The dark matter mini-halo model for HVCs by \citet{Sternberg2002} provides a plausible explanation for some of the HVCs near M31, in particular those at larger projected distances which are isolated in position-velocity space from both the stellar stream and any known dwarf companion of M31. Tidal interaction during their evolution as well as instability and ionisation at larger distances from M31 can explain the small number of HVCs and their proximity to M31. At the same time, HVCs would be sensitive test particles to probe the physical conditions in the circumgalactic environment of large spiral galaxies.

  \section{Origin of the HVCs near M31}
  \label{sect_origin}
  
  As discussed before, our observations seem to favour two possible scenarios for the origin of the HVCs found near M31: They could be tidally stripped gas clouds or primordial, dark-matter-dominated satellites. In this section we will discuss the evidence for these two scenarios and a few other popular hypotheses on the origin of HVCs.
  
  Several of the clouds could be the result of tidal or ram-pressure stripping in connection with the interaction between M31 and its present or former satellite galaxies. This could be the case for the concentration of HVCs (number 1--6 in our catalogue) detected near the south-eastern edge of the \ion{H}{i} disc of M31 which spatially and kinematically overlap with the giant stellar stream \citep{Ibata2001,Ferguson2006}. The nature of this HVC complex was discussed in detail by \citet{Westmeier2005a} based on high-resolution WSRT observations. The HVCs are characterised by a particularly complex velocity structure (see Fig.~\ref{fig_m31fields}) suggestive of turbulent motions. Unfortunately, recent numerical simulations of the stellar stream \citep{Fardal2006,Fardal2007,Mori2008} only focussed on non-collisional particles and did not include gas, so that an association of the HVCs with the stellar stream can not be confirmed independently. Another specific candidate for tidal or ram-pressure stripping is HVC number~14 which is located close to NGC~205. Again, the radial velocity of the HVC is similar to the velocities of the \ion{H}{i} gas detected in NGC~205 (see \citealt{Westmeier2005a} for details).
  
  A tidal origin would naturally explain the observed concentration of HVCs in the vicinity of M31 where tidal forces and halo gas density are increased. Another consequence of the tidal scenario would be moderately high metallicities due to processing of the HVC gas in satellite galaxies and possible mixing with enriched gas from the halo or disc of M31. The only confirmed example of tidally stripped high-velocity gas in the vicinity of the Milky Way is the Magellanic Stream (including the Leading Arm) which shows typical metallicities of about $0.1 \ldots 0.4$ (e.g.,~\citealt{Wakker2001,Sembach2001}). Other large HVC complexes near the Milky Way show comparable metallicities of the order of $0.1$ (see \citealt{Wakker2001} for a comprehensive review), demonstrating that the gas must have been processed and moderately enriched in the past. Direct metallicity measurements of HVCs near M31, however, will be hampered by the lack of suitable background sources for absorption line spectroscopy.
  
  As discussed in Sect.~\ref{sect_cdmcomp}, some of the HVCs around M31 could also be primordial dark-matter haloes left over from structure formation in the Local Group. This could in particular be the case for isolated HVCs at larger projected distances from M31. In this case, we would expect to observe primordial metal abundances unless the gas has been enriched through interaction with halo gas or luminous dark-matter satellites. As mentioned before, another consequence would be the presence of ionised or pure dark-matter satellites out to distances of several hundred kpc from M31 or the Milky Way. Depending on the nature of dark matter, pure dark-matter haloes would be very hard to detect, but they could leave noticeable imprints (e.g., \citealt{Wang2007}).
  
  The discovery of HVCs and several new satellite galaxies near M31, however, cannot directly solve two fundamental problems in connection with CDM structure formation scenarios. First of all, the number of gaseous and luminous satellites would still be one order of magnitude below the predictions of CDM simulations, and we would still have to assume the additional presence of hundreds of pure dark-matter haloes to solve the `missing satellites' problem. Second, The discovery of new satellites does not solve the problem that satellite galaxies around the Milky Way and M31 appear to be arranged in great planes with large inclination angles (e.g., \citealt{Kroupa2005,Koch2006}). Even if the new satellites turned out to have a spherical distribution, the anisotropy of the most massive satellites would remain. This anisotropy has been explained with the preferred infall of dark-matter haloes along the filaments of the cosmic web and subsequent evolution of the satellite orbits within the triaxial potential of their host galaxy's dark-matter halo \citep{Libeskind2005,Zentner2005}. An alternative hypothesis claims that many satellite galaxies around the Milky Way and M31 are tidal dwarf galaxies instead of cosmological dark-matter satellites \citep{Metz2007}.
  
  Another possible explanation for the more isolated HVCs found around M31 is a tidal origin in connection with a previous close encounter between M31 and M33. According to recent test particle simulations by \citet{Bekki2008}, tidal interaction between the two galaxies during their first encounter between $4$ and $8~\mathrm{Ga}$ ago would have stripped gas from M33 which could account for the \ion{H}{i} bridge found by \citet{Braun2004} between M31 and M33. \citet{Bekki2008} suggests that tidally stripped gas from M33 could also have formed some of the HVCs near M31. Consequently, the HVCs would have metallicities similar to those observed in the outer gas disk of M33. The limitations of current observations and simulations, however, do not allow us to further investigate this interesting option.
  
  Another hypothesis suggests that HVCs are the result of galactic outflows due to individual supernovae or the galactic fountain process (e.g., \citealt{Bregman1980,Booth2007}). This hypothesis appears to be inconsistent with the typically low metallicities of Galactic HVCs of the order of 10~per~cent of the solar value (see \citealt{Wakker2001} for a comprehensive review). However, the measured values vary over some range, indicating that the HVCs around the Milky Way do not form a homogeneous population. In addition, metallicities have only been measured for the large HVC complexes near the Milky Way, whereas the metallicities of most CHVCs are unknown. At the same time, we failed to detect the expected large population of CHVCs around M31 in our Effelsberg survey, suggesting that CHVCs could be intrinsically compact objects at small distances from the Milky Way and M31. Hence, CHVCs could be promising candidates for gas ejected by galactic outflows.
  
  \section{Summary and conclusions}
  \label{sect_summary}
  
  We used the 100-m radio telescope at Effelsberg to map a large area around the Andromeda Galaxy, M31, in the 21-cm line emission of neutral atomic hydrogen. Our survey extends out to a projected distance of about $140~\mathrm{kpc}$ in the south-eastern direction (equivalent to about two thirds of the projected distance towards M33) and about $70~\mathrm{kpc}$ in the north-western direction. With this map layout we are able to fill the previously existing gap between the outer boundary of the GBT survey of \citet{Thilker2004} at about $50~\mathrm{kpc}$ projected distance from M31 and the upper limit of about $150~\mathrm{kpc}$ for the distance of HVCs as derived from the non-detections in nearby galaxy groups \citep{Zwaan2001,Braun2001,Pisano2004}. The achieved spectral baseline RMS is $45~\mathrm{mK}$ at $2.6~\mathrm{km \, s}^{-2}$ velocity resolution, corresponding to a $3 \, \sigma$ \ion{H}{i} column density detection limit of $2.2 \times 10^{18}~\mathrm{cm}^{-2}$ for the warm neutral medium ($\Delta v = 25~\mathrm{km \, s}^{-1}$ FWHM). This translates into an \ion{H}{i} mass sensitivity of $8 \times 10^{4}~{\rm M}_{\odot}$.
  
  In total, we detected 17~individual HVCs and several regions of more extended extra-planar gas all around M31. The discrete clouds are predominantly unresolved by the HPBW of $9~\mathrm{arcmin}$ and characterised by typical \ion{H}{i} masses of a few times $10^{5}~{\rm M}_{\odot}$. We did not detect any clouds beyond a projected distance of about $50~\mathrm{kpc}$, suggesting that HVCs are generally found in proximity of their host galaxies. In particular, we did not find an extended populations of hundreds of CHVCs as observed around the Milky Way, suggesting that the Galactic CHVCs are intrinsically small clouds in the immediate vicinity of the Milky Way.
  
  A comparison with the Local Group population model of CHVCs, as proposed by \citet{deHeij2002b}, reveals that their best-fitting model {\#}9 is discarded by our data with high confidence. Neither the observed projected radial distribution nor the \ion{H}{i} mass function of HVCs around M31 can be explained by the model. Instead, we find that a Gaussian radial scale length of the order of $50~\mathrm{kpc}$ can best explain the observed projected distribution of HVCs around M31. In addition, we find that the HVCs are also distinct from the M31 satellite galaxies through typically lower \ion{H}{i} masses and smaller projected distances from M31.
  
  CDM-based structure formation simulations by \citet{Kravtsov2004} suggest that about 50 to 100~dark-matter haloes with total gas masses of greater than $10^{6}~{\rm M}_{\odot}$ should exist within about $300~\mathrm{kpc}$ of M31. Only~2 to~5 of these haloes should be located within $50~\mathrm{kpc}$ of M31, suggesting that some of the HVCs found near M31 could instead be tidally stripped gas from present or former satellite galaxies of M31. This idea is supported by our high-resolution follow-up observations of several HVCs \citep{Westmeier2005a}.
  
  The lack of detections beyond $50~\mathrm{kpc}$ projected radius, however, is in conflict with the predictions made by \citet{Kravtsov2004}. A possible explanation of this discrepancy could be ionisation as a result of decreasing pressure of the ambient coronal medium at larger distances from M31, as suggested by hydrostatic simulations of \citet{Sternberg2002}. An important consequence of this scenario would be the presence of hundreds of mainly ionised or pure dark-matter satellites near large spiral galaxies, such as the Milky Way and M31, which would be undetectable in the 21-cm line of neutral hydrogen. Finding these almost invisible satellites would be an important observational result in support of CDM cosmologies.
  
  Another promising scenario recently discussed by \citet{Bekki2008} is a possible tidal interaction between M31 and M33 during their previous encounter. This would have stripped gas from M33 which could have formed some of the HVCs observed near M31. More detailed and extensive observations and simulations will be required to further investigate this interesting scenario.
  
  With our Effelsberg \ion{H}{i} survey of M31 we have shown that HVCs are most likely concentrated around the large spiral galaxies with typical distances of no more than a few $10~\mathrm{kpc}$. It is likely that different physical processes, such as tidal stripping, accretion of primordial dark-matter haloes, or galactic outflows, have contributed to the HVC populations of M31 and the Milky Way. The general volatility of these processes could naturally explain some of the differences between the HVC populations of M31 and the Milky Way, for example the extended and complex gaseous streams of the Magellanic Clouds which have no counterpart in M31.
  
  \section*{Acknowledgments}
  
  This project was supported by the German Research Foundation (DFG) through grants KE757/4--1 and 4--2. Based on observations with the 100-m telescope of the MPIfR (Max-Planck-Institut f{\"u}r Radioastronomie) at
Effelsberg.
  
  \bibliographystyle{mn2e}
  \bibliography{westmeier.bib}
  
  \bsp
  
  \label{lastpage}
  
\end{document}